\documentclass[12pt,reqno]{amsart}
\usepackage{amsmath,amssymb,amsfonts,amsthm}
\usepackage[mathscr]{eucal}
\usepackage[all]{xy}
\usepackage{hyperref}
\title[BRST analysis of general mechanical systems]
{BRST analysis of general mechanical systems}

\author{D.S. Kaparulin,  S.L.  Lyakhovich and A.A. Sharapov}

\address{Department of Quantum Field Theory, Tomsk State University, Lenin ave. 36, Tomsk 634050, Russia.}

\email{dsc@phys.tsu.ru, sll@phys.tsu.ru, sharapov@phys.tsu.ru}

\thanks{The work is done partially under the project 2.3684.2011 of Tomsk State
University, FTP contract  14.B37.21.0911 and the RFBR grant
13-02-00551. DSK and AAS appreciate the financial support from
Dynasty Foundation, SLL acknowledges support from the RFBR grant
11-01-00830-a.}

\theoremstyle{remark}

\renewcommand{\simeq}{\cong}

\begin{document}

\maketitle

\begin{abstract} We study the groups of local BRST cohomology
associated to the general systems of ordinary differential
equations, not necessarily Lagrangian or Hamiltonian. Starting
with the involutive normal form of the equations,
we explicitly compute certain cohomology groups having clear physical meaning.
These include the groups of global symmetries, conservation laws and Lagrange structures.
It is shown that the space of integrable Lagrange structures is
naturally isomorphic to the space of weak Poisson brackets. The
last fact allows one to establish a direct link between the
path-integral quantization of general not necessarily variational dynamics
by means of Lagrange structures and the deformation quantization
of weak Poisson brackets.
\end{abstract}

\section{Introduction}

The BRST methods initially appeared as a uniform tool for quantizing
either Lagrangian gauge theories or  Hamiltonian constrained
dynamics (for review see \cite{HT}). Correspondingly, the two
frameworks have been worked out. The first one, most frequently
referred to as a BV or field-anti-field BRST formalism was
originally aimed at the problem of covariant path-integral
quantization of Lagrangian theories. The second one, commonly called
either the BFV formalism or Hamiltonian BRST formalism is most
suitable for operator or deformation quantization of the Hamiltonian
dynamics. Later on, the BRST formalisms have begun gaining
applications in various problems well beyond the original issue of
quantization, e.g. in  topological field theory \cite{AKSZ}. Though
the BV and BFV methods share basic principles, they use different
prerequisites for constructing the BRST complex and technically are
quite different. The relationship between these approaches was
established in several ways (see e.g. \cite{BF}, \cite{BH},
\cite{GD}).

In the recent years, the BRST methods have been extended beyond the
scope of  Lagrangian or Hamiltonian dynamics \cite{LS0},
\cite{KazLS}. In particular, it was shown that the classical BRST
complex can be systematically constructed for a general dynamical
system, not necessarily Lagrangian or Hamiltonian. If the dynamical
system admits an extra structure, called the \textit{weak Poisson
bracket}, then a consistent deformation quantization can be
performed in the absence of gauge anomalies \cite{LS0}, \cite{CaFe}.
This method can be viewed as a far-reaching extension of the BFV
formalism to not necessarily Hamiltonian dynamics. As a prerequisite
for the deformation quantisation \cite{LS0}, \cite{CaFe}, the
dynamics should be brought to the \textit{involutive normal form}.
This does not restrict the generality, as any regular gauge dynamics
can be equivalently formulated in this way \cite{LS}. For
variational dynamics, the involutive normal form reduces to Dirac's
constrained Hamiltonian system, and the corresponding BRST complex \cite{LS0}
reduces to the BFV one.

The corresponding extension of the BV formalism is relied on the new
concept of a \textit{Lagrange structure}. Existence of the Lagrange
structure is less restrictive for the dynamics than the requirement
for the equations to follow from the variational principle. Given a
Lagrange structure compatible with equations of motion, the
classical theory can be path-integral quantized in several ways
\cite{KazLS}, \cite{LS1}, \cite{LS2}.

Within the BRST approach, the most of the information about the
structure of gauge dynamics is encoded in the groups of local BRST
cohomology.  In particular, the physical observables, global
symmetries, conservation laws, Lagrange structures, quantum
anomalies, consistent interactions and counterterms  are all the
elements of the corresponding cohomology groups. This explains the
paramount role that the concept of local BRST cohomology plays in
the modern quantum field theory.  For Lagrangian theories, several
important general theorems on the structure of local BRST cohomology
groups were obtained in the last decades of XX century. A
comprehensible review of these results can be found in \cite{BBH1},
\cite{BBH2} \cite{BBH}. Recently \cite{KLS2}, some of these general
theorems were systematically extended beyond the class of Lagrangian
dynamics, including the cohomological formulation of Noether's first
theorem \cite{K-S}, \cite{KLS1}. It is not surprising that the local
BRST cohomology groups, being so informative, are not easy to
compute for any nontrivial model, even in Lagrangian setting. More
or less complete description of the groups was obtained only for the
theories of Yang-Mills type \cite{BBH2}, \cite{BBH}. Some groups
have been recently described for the AKSZ-type models in \cite{BG1}.
The BRST cohomology in the case of usual Hamiltonian mechanics was
first considered in \cite{HT1988}.

The present paper is devoted to the study of the groups of local
BRST cohomology for mechanical systems whose dynamics are governed
by ordinary differential equations (ODEs) of general form. In
particular, we do not assume the equations of motion to come from
the least action principle that would impose a strong restriction
on the structure of dynamics. Since the general theory of ODEs is
much more elaborated nowadays than that of PDEs it is reasonable
to expect that the corresponding groups of local BRST cohomology
are more traceable from the standpoint of computability and
physical interpretation. This expectation is generally confirmed
by our results below. The main advantage of working with ODEs is
the existence of an involutive normal form to which any equations
can be locally brought to by introducing auxiliary variables
\cite{LS}. The procedure of bringing the general dynamics
to the involutive form does not impose any restrictions on dynamics,
besides some regularity conditions.
For variational systems, the procedure of passing to
the involutive normal form reduces to the Dirac-Bergmann algorithm
of bringing the general Lagrangian dynamics to the Hamiltonian
system with first and second class constraints.

An important advantage of utilizing the involutive normal form is
that the BRST charge turns out a local functional on the extended
space of trajectories with the integrand involving no more than
the first derivatives of dependent variables. It is the absence of
higher derivatives which gives an efficient control over the
structure of local BRST cohomology and which allows one to bring
the calculations of the most interesting groups to the very end.
In particular, we give a detailed description for the space of
Lagrange structures, which appears to be naturally isomorphic to
the space of weak Poisson structures associated with an involutive
system of ODEs. In the other words, each Lagrange structure
defines a weak Poisson bracket and vice versa. This new fact is
important for linking two different branches of the BRST
formalism, the BV and BFV ones, in the more general class of
dynamics than variational. In particular, it allows one to bridge
two different approaches to the quantization of (non-)Lagrangian
gauge systems: the path-integral quantization by means of Lagrange
structure  and the deformation quantization  of weak Poisson
structure. Although our consideration is restricted to the systems
of ODEs, the most of results and computational technique   can
hopefully be transferred to the field-theoretical models governed
by PDEs of evolutionary type. This can require, in principle, a
due account of space locality that we do not address in this work.
In covariant field theories, however, the space locality is
usually related to the locality in time. That is why we can hope
that our results on the local BRST cohomology groups, being
derived for the systems local in time, will avoid obstructions
related to the pure spacial non-locality, at least in the
covariant field theories.

Let us also mention some of possible applications of the BRST
analysis in the optimal control theory, where the gauge freedom is
reinterpreted as the degree of controllability (for an extended
discussion see \cite{LS}, \cite{AS}). Among various issues
considered in the optimal control there are those concerning
isomorphisms of controllable systems and normal forms to which a
given controllable system can be brought to by a suitable
transformation (static or dynamical feedback equivalence,
Lie-B\"acklund isomorphisms, flatness, etc.). The groups of local
BRST cohomology, being invariants of all such transformations,
provide an efficient tools for attacking these problems.
Specifically, one can hope to use them as the spaces of obstructions
to global equivalence between two controllable systems and/or as the
invariant characterization of normal forms.

The paper is organized as follows. In the next section, we recall
the definition of the involutive normal form for a general system of
ODEs. For this normal form, we explicitly identify the generators of
gauge symmetries and the Noether identities, which are necessary
inputs for constructing the classical BRST charge. The classical
BRST complex is discussed in Sec. 3. Here we first define the
extended symplectic space of trajectories endowed with the
Hamiltonian action of the classical BRST charge. We briefly comment
on the structure of the classical BRST differential and explain the
physical meaning of simplest BRST cohomology groups. Sec. 4 is
devoted to computation of the local BRST cohomology groups both in
the spaces of local functions and functionals. The computation
follows certain systematic procedure. It utilizes a special
filtration in the infinite jet spaces that are respected by the
Koszul-Tate and longitudinal differentials. This makes possible to
work exclusively with functions on finite dimensional spaces and
define the corresponding cohomology groups as direct limits.
Besides, we intensively exploit the long exact sequences in
cohomology (which in our exposition look like exact triangles) and
the mapping cone construction. In Sec. 5, we review the construction
of the total BRST charge, which is a basic ingredient of the
path-integral quantization of (non-)Lagrangian gauge systems. The
total BRST charge is defined as a deformation of the classical BRST
charge and then reinterpreted  as a $L_\infty$-algebra on a certain
space of functionals with the first structure map given by the
classical BRST differential. We show that the total BRST charge  of
an involutive mechanical systems is completely specified by the
first and second structure maps. It is the second structure map
(weak anti-bracket) which is identified with an integrable Lagrange
structure. Sec. 6 establishes a one-to-one correspondence between
the spaces of integrable Lagrange structures and weak Hamiltonian
structures associated with involutive systems of ODEs. The proof of
the correspondence is relied on the results of Sec. 4. In
the final section, we briefly review the BRST formulation of the
weak Hamiltonian structures in terms of two generating functions
proposed in \cite{LS0}. Then, using a superfield approach,  we
present a systematic procedure for explicit construction of the
total BRST charge from the two generating functions of a weak
Hamiltonian structure.

\section{Involutive systems of ODEs}

In this paper, we consider autonomous systems of ordinary
differential equations in the so-called \textit{involutive normal
form}:
\begin{equation}\label{INF}
    \dot{x}^{i}+V^{i}(x)+\lambda^{\alpha}R_{\alpha}^{i}(x)=0\,,\qquad
    T_{a}(x)=0\,.
\end{equation}
Here the dot over $x$'s stands for the derivative in the
independent variable $t$, called the ``time'', and the dependent
variables $x$'s and $\lambda$'s are treated as local coordinates
on the phase space of the system. To avoid topological
complications we shall assume the phase space to be the linear
manifold $\mathbb{R}^{n+m}=\mathbb{R}^n\times \mathbb{R}^m$ with
the global coordinates $\{x^i,\lambda^\alpha\}$. The vector field
$V$ entering the differential equations is called the
\textit{drift}, while the collection of the vector fields
$R=\{R_\alpha\}$ is referred to as the \textit{gauge
distribution}. The algebraic equations defined by the functions
$T_a(x)$, $a=1,\ldots, l$, are called the \textit{constraints}.
Involutivity implies that the following identities hold with some
structure functions $A$, $B$, $D$, $E$ and vector fields $C$, $F$:
\begin{equation}\label{INV}
\begin{array}{ll}
   [R_\alpha, T_a]= A_{\alpha a}^bT_b\,,&  [R_\alpha, R_\beta]=B_{\alpha\beta}^\gamma R_\gamma -T_a C^{a}_{\alpha \beta}
   \,,\\[3mm]
  [V,T_a]= D_a^bT_b \,,\qquad & [V,R_\alpha]=E_\alpha^\beta R_\beta-T_aF_\alpha^a
  \,,
\end{array}
\end{equation}
Hereafter the square brackets denote the Schouten bracket in the
space $\Lambda(\mathbb{R}^n)=\bigoplus_{k=0}^n
\Lambda^k(\mathbb{R}^n)$ of smooth polyvector fields on
$\mathbb{R}^n$. In particular, the bracket of a vector field $v$
with a function $f$ ($0$-vector field)  is understood as the Lie
derivative of the function along the vector field, $[v,f]={L}_v f$
and the Schouten bracket of two vector fields is given by their
commutator\footnote{For the general definition of the Schoten
bracket see formula (\ref{SB}) below.}.

Let $\Sigma$ denote the zero locus of the constraints, i.e.,
$\Sigma=\{x\in \mathbb{R}^n\, | \, T_a(x)=0, \, a=1, \ldots, l
\}$. In what follows we assume the variety $\Sigma$ to be
nonempty, the constraints $T_a$ to be functionally independent and
the vector fields $R_\alpha$ to be linearly independent at each
point of $\Sigma$. This amounts to the full rank condition for
appropriate matrices, namely,
\begin{equation}\label{IRRED}
    \mathrm{rank} \big(\partial_iT_a(x)\big)=
    l\,,\qquad \mathrm{rank} \big(R_\alpha^i(x)\big)=m\,,\qquad \forall x\in \Sigma\,.
\end{equation}
The first equality also ensures that $\Sigma \subset \mathbb{R}^n$
is a smooth submanifold. It is called the \textit{constraint
surface}.

The property of the system (\ref{INF}) ``to be involutive'', being
defined by (\ref{INV}), actually captures two different aspects,
which should not be mixed up. To discuss either of them, let us
first introduce the exterior ideal $I\subset
\Lambda(\mathbb{R}^n)$ generated by the $1$- and $0$-vector fields
$V$, $R$'s, and $T$'s that determine the system (\ref{INF}).
Relations (\ref{INV}) mean that $I$ is closed for the Schouten
bracket, $[I,I]\subset I$, and hence $I$ is a subalgebra of the
graded Lie algebra $\Lambda(\mathbb{R}^n)$. From the geometrical
viewpoint, this means that the gauge distribution $R$ is tangent
to and integrable on $\Sigma$ and it remains to be so even when
completed by the drift $V$. Furthermore, the restriction
$R|_{\Sigma}$ of the gauge distribution to the constraint surface
is invariant under the action of $V|_{\Sigma}$. The distribution
$R|_{\Sigma}$, being integrable and having a constant rank, defines a
regular foliation $\mathcal{F}$ on $\Sigma$. The leaves of
$\mathcal{F}$ are called the \textit{gauge orbits}. The space of
leaves $M=\Sigma/\mathcal{F}$ is known as the \textit{physical
phase space}. Notice that
 $\dim M=n-l$ whenever $M$ is a Housdorff manifold. The terminology ``gauge
distribution'' and ``gauge orbits'' is justified by the fact that
the system (\ref{INF}) enjoys infintesimal gauge symmetries of the form
\begin{equation}\label{GSYM}
    \delta_{\varepsilon} x^i=-\varepsilon^\alpha
    R_\alpha^i(x)\,,\qquad
    \delta_{\varepsilon}\lambda^\alpha=\dot\varepsilon{}^\alpha+\varepsilon^\beta(E_\beta^\alpha(x)+\lambda^\gamma
    B_{\beta\gamma}^\alpha(x))\,,
\end{equation}
$\varepsilon$'s being infinitesimal gauge parameters.   As a
result, the system of equations (\ref{INF}) is underdetermined and
one can choose $\lambda$'s to be arbitrary functions of time. It
is easy to see \cite{LS} that  any gauge invariant $t$-local value
associated with the phase space $\mathbb{R}^{n+m}$ can be
represented by a function on $\Sigma$ which is constant along each
gauge orbit. Therefore, the algebra of physical observables is
isomorphic to the commutative algebra of functions on the physical
phase space $M$. In the next sections, we shall rediscover  and
reinterpret the last fact within a cohomological analysis. The
time evolution of the physical observables is generated by the
drift $V$, or more precisely, by its projection on  $M$. It is the
involutivity of the distribution $R|_{\Sigma}$ that was the main
reason in \cite{LS} to call the normal form (\ref{INF})
involutive.

Also, there is another reason to use the term  ``involutive''. It is
related to a general notion of involution for the system of
differential algebraic equations \cite{Seiler}. Loosely, a system is
said to be involutive if it contains no implicit integrability
conditions. For the systems of the form (\ref{INF}) these hidden
integrability conditions may appear when one differentiates the
constraints with respect to $t$ and eliminates then the velocities
$\dot x$ with the help of the differential equations. In general,
this can result in new algebraic constraints on the phase-space
variables. Adding these new constraints to the original ones and
extracting functionally independent among them, one can repeat the
above procedure once and again producing further integrability
conditions.  This is known as the completion of a system to
involution.  Taking the total derivative of the constraints $T_a$
and making use of relations (\ref{INV}), we find
\begin{equation}\label{NI}
\frac{ d T_{a}}{dt}=\frac{\partial T_{a}}{\partial x^{i}}\left(
\dot{x}^{i}+V^{i}+\lambda^{\alpha}R_{\alpha}^{i}\right)-(\lambda^{\alpha}A_{\alpha
a}^{b}+D_{a}^{b})T_{b}\,.
\end{equation}
So, the time evolution  preserves  $\Sigma$ and no new constraints
on $x$'s or $\lambda$'s arise. In other words, the system
(\ref{INF}) is involutive provided that the first and third
conditions in (\ref{INV}) are satisfied.  Notice that the absence of
hidden integrability conditions implies simultaneously the presence
of the \textit{Noether identities} (\ref{NI}) among the equations of
motion (\ref{INF}). Indeed, the total derivative of  every algebraic
equation (i.e., its differential consequence)  has to be given by a
linear combination of the algebraic and differential equations.

If the constraints $T_a(x)$ are chosen to be independent, then no
other Noether identities can exist. Although the Lie closedness of
the exterior ideal $I$ ensures the involutivity of the system in the
sense  of the absence of integrability conditions, the converse is
not true. In particular, the presence of the Noether identities
(\ref{NI}) has nothing to do with involutivity of the distribution
$R|_{\Sigma}$. When the latter is not involutive, the system
(\ref{INF}) is still underdetermined, but the corresponding gauge
transformations involve higher derivatives of the gauge parameters
$\varepsilon^\alpha$, so that $\dim M< n-l$. The last situation is
typical for the so-called affine control systems \cite{AS}.

Due to the full rank conditions (\ref{IRRED}), both the gauge
symmetry transformations (\ref{GSYM}) and the Noether identities
(\ref{NI}) are \textit{irreducible} in the usual sense \cite{HT}.

In \cite{LS}, it was shown that \textit{any} system of ODEs can be
locally brought to an involutive normal form
(\ref{INF}),(\ref{INV}) by introducing auxiliary variables. The
differential algebraic equations (\ref{INF}) with the structure
functions subject to the involutivity conditions (\ref{INV}) can
thus be taken as a starting-point for the general theoretical
analysis of local dynamics governed by ODEs. Equations (\ref{INF})
can also be regarded as a generalization of the Dirac-Bergmann
normal form known in the constrained Hamiltonian dynamics
\cite{HT}. In the Hamiltonian situation, the algebraic equations
$T_a=0$ constitute the set of all the first and second class
constraints (both primary and secondary), the variables
$\lambda^\alpha$ correspond to the Lagrange multipliers to the
first class constraints, whose Hamiltonian vector fields are
identified with the generators $R_\alpha$ of the gauge
distribution. Finally, the Poisson bracket of the Hamiltonian
generates the drift $V$. Upon these identifications, the
involutivity conditions (\ref{INV}) are equivalent to completeness
of the set of Hamiltonian constraints.

The advantage of the involutive normal form over the other
equivalent representations of ODEs is the simple structure of the
gauge transformation (\ref{GSYM}) and the Nother identities
(\ref{NI}), namely, the absence of higher derivatives. This will
allow us to perform an exhaustive cohomological analysis of the
system and give an explicit description for all the relevant
groups of local BRST cohomology.

\section{Local BRST complex}
Within the BRST formalism the equations of motion (\ref{INF}), the
gauge transformations (\ref{GSYM}) and the Noether identities
(\ref{NI}) are all incorporated in a singe object $\Omega_1$ called
the \textit{classical BRST charge}. The construction of $\Omega_1$
is made by the homological perturbation theory and it works, in
principle, for arbitrary systems of PDEs. Referring to \cite{KazLS}
and \cite{KLS2} for details, here we just present the ``cookbook
recipe'' for the system at hand. First, the space $\mathbb{R}^{n+m}$
of the original variables $x$'s and $\lambda$'s is extended by the
new variables $\eta_i$, $\eta^a$, $c^\alpha$, and $\xi_a$ usually
called the \textit{ghosts}. The number of $\eta$'s, $c$'s, and
$\xi$'s coincides, respectively, with the number of equations of
motion, gauge symmetries, and Noether identities. It is convenient
to introduce the collective notation $\varphi^I=\{x^i,
\lambda^\alpha, \eta_i, \eta^a, c^\alpha, \xi_a\}$.  At the next
step the collection $\varphi$ redoubles by adding the dual variables
$\bar{\varphi}_J=\{\bar x_{i},\bar\lambda_\alpha, \bar \eta^i,\bar
\eta_a, \bar c_{\alpha},\bar \xi^a\}$ called the momenta. The
variables from the either collection are considered to be arbitrary
functions of time. Introducing the canonical Poisson bracket
\begin{equation}\label{PB}
    \{\varphi^I(t),\varphi^J(t')\}=0\,,\qquad \{\bar \varphi_I(t),\varphi^J(t')\}=\delta_I^J\delta(t-t')\,,
    \qquad\{\bar \varphi_I(t),\bar \varphi_J(t')\}=0\,,
\end{equation}
one can think of $\varphi^I(t)$ and $\bar \varphi_J(t)$ as
coordinates on an infinite-dimensional phase-space $V$. For an
obvious reason we shall call the points of $V$
\textit{trajectories}. The space $V$ is actually a multigraded
superspace. The gradings are defined by prescribing the following
degrees to the dependent variables:
$$
\begin{array}{c}
\mathrm{gh}(x^i)=\mathrm{gh}(\lambda^\alpha)=0\,,\qquad
\mathrm{gh}(\eta_i)=\mathrm{gh}(\eta^a)=-1\,,\qquad
\mathrm{gh}(c^\alpha)=1\,,\qquad \mathrm{gh}(\xi_a)=-2\,,\\[3mm]
\mathrm{gh}(\bar\varphi_J)=-\mathrm{gh}(\varphi^J),\\[3mm]
\epsilon(\varphi^I)=\mathrm{gh}(\varphi^I)\,,\quad \epsilon(\bar\varphi_J)=\mathrm{gh}(\bar\varphi_J)\quad (\mathrm{mod} \;2)\,,\\[3mm]
\mathrm{deg}(x^i)=\mathrm{deg}(\lambda^\alpha)=\mathrm{deg}
(\bar\eta_i)=\mathrm{deg}(\bar\eta^a)=\mathrm{deg}(c^\alpha)=\mathrm{deg}(\bar\xi^a)=0\,,\\[3mm]
\mathrm{deg}(\bar{x}_i)=\mathrm{deg}(\bar{\lambda}_\alpha)=\mathrm{deg}(\eta^i)=\mathrm{deg}(\eta_a)=1\,,\qquad
\mathrm{deg}(\bar c^\alpha)=\mathrm{deg}(\xi_a)=2\,,\\[3mm]
\mathrm{Deg}(\varphi^I)=0\,,\qquad
\mathrm{Deg}(\bar\varphi_J)=1\,.
\end{array}
$$
The $\mathbb{Z}$-grading defined by the first and second lines is
known as the \textit{ghost number}. As is seen the ghost numbers
of momenta are opposite to the ghost numbers of the ``position
coordinates''. Since we are dealing with a mechanical system
without fermionic degrees of freedom, the Grassmann parity
$\epsilon\in \mathbb{Z}_2$ of all the variables is uniquely
determined by their ghost number (the third line). Besides, there
are two auxiliary $\mathbb{N}$-gradings: the \textit{resolution
degree} and the \textit{momentum degree} denoted respectively by
$\mathrm{deg}$ and $\mathrm{Deg}$. The former is crucial for the
homological perturbation theory (hence the name), while the latter
just counts the number of momenta in homogenous expressions.

Let $\Phi^A=(\varphi^I, \bar\varphi_J)$ denote the whole set of
coordinates on the infinite-dimensional phase-space of
trajectories $V$. By a \textit{local function} on $V$ we mean a
function $f(t)$ that depends on time through the trajectory
$\Phi^A(t)$ and its $t$-derivatives up to some finite order, that
is, $ f=f(\Phi, \dot \Phi,\ldots, \stackrel{_{(k)}}{\Phi}{})$. The
local functions form a graded supercommutative algebra with
respect to the point-wise multiplication,  which we denote by $F$.
\textit{Local functionals} on $V$ are by definition integrals of
local functions over closed intervals $I\subset \mathbb{R}$. In
the sequel we shall assume the integration domain $I$ to be fixed
once and for all. Then each local functional $\int_I f dt$ is
completely specified by its integrand, i.e., by the local function
$f\in F$. The correspondence between local functionals and
functions is not one-to-one. Indeed, if the boundary conditions
for the admissible trajectories of $V$ are chosen in such a way
that $g|_{\partial I}=0$ for some $g\in F$, then $\int_I
(Dg)dt=0$, where $D=d/dt$ is the operator of total time
derivative. To eliminate this ambiguity we impose an equivalence
relation, whereby two local functionals are considered as
equivalent if they only differ by boundary terms. In other words,
the variational derivatives of equivalent functionals coincide.
Then two local functions $f$ and $f'$ determine equivalent
functionals iff $f'-f =Dg$ for some $g\in F$. This allows us to
identify the equivalence classes of local functionals with the
quotient $F/DF$ of the space of local functions by the subspace of
total time derivatives\footnote{This equivalence relation on the
space of local functionals is not so artificial as might appear at
first sight. In actual fact it is customary to impose zero
boundary conditions on all the ghosts and momenta as well as their
derivatives. Then the integral of the total derivative of a local
function with nonzero ghost number or momentum degree is equal to
zero automatically. Another situation where the equivalence
relation above establishes an isomorphism between the local
functions and functionals is the case of differential equations on
circle, $I=S^1$.}. Notice that the latter subspace is not a
subalgebra in $F$.

The classical BRST charge $\Omega_1$ is now defined to be a local
functional satisfying the following set of conditions:
\begin{enumerate}
    \item $\epsilon(\Omega_1)=1$, $\mathrm{gh}(\Omega_1)=1$,
    $\mathrm{Deg}(\Omega_1)=1$;
    \item $\Omega_1 =\displaystyle\int dt\Big[
\bar{\eta}_{i}(\dot{x}^{i}+V^{i}+\lambda^{\alpha}R_{\alpha}^{i})+\bar{\eta}^{a}T_{a}$
    \begin{equation}\label{BOUND}
    \begin{array}{rcl}
%\\[3mm]
&&+c^{\alpha}\big(\dot{\bar{\lambda}}_{\alpha}+R_{\alpha}^{i}\bar{x}_{i}+
\bar\lambda{}_\beta
E^\beta_\alpha+\lambda^{\beta}B_{\beta\alpha}^{\gamma}\bar{\lambda}_{\gamma}+\lambda^{\beta}C_{\alpha\beta}
^{ai}\bar{\eta}_{i}\eta_{a}-F_{\alpha}^{ai}\bar{\eta}_{i}\eta_{a}-A_{\alpha
a}^{b}\bar{\eta}^{a}\eta_{b}-\frac{\partial
R_{\alpha}^{i}}{\partial x^{j}}\bar{\eta}_{i}\eta^{j}\big)\\[3mm]
&&-\bar{\xi}^{a}\big( \eta_{b}D^{b}_{a}+\lambda^{\alpha}A_{\alpha
a}^{b}\eta_{b}-\eta^{i}\frac{\partial T_{a}}{\partial
x^{i}}+\dot{\eta}_{a}\big)\Big]+\cdots\,,
\end{array}
\end{equation}
where the dots stand for the terms at least quadratic in $c$'s and
$\bar\xi$'s;
\item $\{\Omega_1,\Omega_1\}=0$.
\end{enumerate}
The first condition defines $\Omega_1$ to be an odd functional of
ghost number 1 with linear dependence of momenta. (The subscript $1$
in the notation $\Omega_1$ just points to the linear dependence of
momenta.) The second condition defines the leading terms in the
expansion of $\Omega_1$ according to the resolution degree. The
higher order terms are determined from the \textit{classical master
equation} (3) by means of the homological perturbation theory
\cite{HT}. On this account one can regard (2) as a ``boundary
condition'' for the master equation (3). Notice that the vanishing
of the Poisson square of $\Omega_1$ is a nontrivial condition to
satisfy as the functional $\Omega_1$ is odd. A general theorem
proved in \cite{KLS2} ensures that the classical BRST charge always
exists and is unique up to a canonical transformation in $V$.

The Hamiltonian vector field
\begin{equation}\label{BRST-DIFFERENTIAL}
s_0=\{\Omega_1, \;\cdot\;\}=\delta+\gamma + \sum_{r=1}^\infty
    s_0^{_{(r)}}\,,\\[4mm]
\end{equation}
generated by the classical BRST charge is called the
\textit{classical BRST differential}. In the expression above it
is expanded according to the resolution degree such that
$$
\deg \,\delta=-1\,,\qquad \deg \,\gamma=0\,,\qquad \deg\,
    s_0^{_{(r)}}=r\,.
$$
Clearly, the action of $s_0$ differentiates the algebra of local
functions. The vector fields $\delta$ and $\gamma$ are known as the
\textit{Koszul-Tate differential} and the \textit{longitudinal
differential}, respectively \cite{HT}. Since $s_0^2=0$ and
$\mathrm{gh}\, s_0=1$, the classical BRST differential makes the
algebra $F$ into a cochain complex with respect to the ghost number.
Considering that $\mathrm{Deg}\, s_0=0$, the complex $F$ splits into
the direct sum of complexes with definite momentum degree. We denote
the corresponding cohomology groups by $H^g_m(s_0)$; here the
superscript refers to the ghost number, while the subscript points
on the momentum degree. Since the action of the variational vector
field $s_0$ commutes with the time derivative, we have the short
exact sequence of complexes
$$
\xymatrix{0 \ar[r]&{DF}\ar[r]^-{i}&{F}\ar[r]^-{p} & F/DF \ar[r]&0}
$$
where $i$ is the natural inclusion and $p$ is the canonical
projection. As we have explained above the quotient $F/DF$ is
naturally identified with the space of local functionals on $V$.
Let us denote its cohomology groups by $H^g_m(s_0|D)$. We shall
refer to $H^g_m(s_0|D)$ as the groups of \textit{relative BRST
cohomology} or cohomology of $s_0$ modulo $D$. The classes of
relative BRST cohomology are given by the equivalence classes
$f+s_0F+DF$ where $f\in F$ and $s_0 f\in DF$.

The identity $s_0^2=0$, being expanded with respect to the
resolution degree, implies the infinite sequence of equalities
\begin{equation}\label{s2}
    \delta^2=0\,,\qquad [\delta,\gamma]=0\,,\qquad \gamma^2=-[\delta,
    s_0^{_{(1)}}]\,, \qquad \ldots
\end{equation}
As is seen, the Koszul-Tate differential squares to zero by itself
defining thus one more coboundary operator in $F$ and $F/DF$. Let
us write $H^g_m(\delta)$ and $H^g_m(\delta|D)$ for the
corresponding cohomology groups. Then the second and third
relations in (\ref{s2}) suggest that the longitudinal differential
$\gamma$ induces a coboundary operator in the $\delta$-cohomology:
if $[f]\in H(\delta)$, then we set $\gamma ([f])=[\gamma f]$. (By
abuse of notation we denote this induced coboundary operator  by
the same letter $\gamma$.) A similar definition applies to the
relative $\delta$-cohomology making the space $H(\delta|D)$ into a
cochain complex with respect to $\gamma$. We let $H(\gamma,
H(\delta))$ and $H(\gamma, H(\delta|D))$ denote the corresponding
cohomology groups. Besides the momentum degree, these
$\gamma$-cohomology groups are also graded by the resolution
degree as $\deg\,\gamma=0$.

\vspace{3mm}\noindent \textit{Remark}. Since the classical BRST
charge $\Omega_1$ is linear in momenta, the Hamiltonian action of
$s_0$ is completely determined by its restriction on local functions
with zero momentum degree. This restriction defines a homological
vector field on the $\varphi$-space, which is also called the
classical BRST differential \cite{KazLS}. It is the terms of
resolution degree $-1$ and $0$ of this last homological vector field
that are usually referred to as the Koszul-Tate and longitudinal
differentials. Geometrically, one can think of $s_0$ as a canonical
lift (the Lie derivative construction) of the homological vector
field from the space of $\varphi$-trajectories to its cotangent
bundle $V$.

\vspace{3mm}

Notice that the Koszul-Tate differential $\delta$ decreases the
resolution degree exactly by one unit in contrast to $s_0$, which
is inhomogeneous.  This allows us to interpret $F$ and $F/DF$ as
the \textit{chain} complexes with respect to the resolution
degree. The corresponding homology groups in degree $r$ will be
denoted by $H^{{(r)}}(\delta)=\bigoplus H_m^{{(r)}}(\delta)$ and
$H^{{(r)}}(\delta|D)=\bigoplus H_m^{{(r)}}(\delta|D)$. (We enclose
the superscript in round brackets to distinguish it from the ghost
number. The lower index indicates the momentum degree as before.)
This change-over from the $\delta$-cohomology to the
$\delta$-homology and vice versa is very helpful for formulating
and proving various assertions  below.

The local BRST cohomology of regular systems of PDEs was
systematically studied in our recent paper \cite{KLS2}. Being
applied to ODEs, the results of \cite{KLS2} lead to the conclusion that all  the
nontrivial BRST groups  concentrate in resolution degrees 0 and 1
and are given by
$$
\begin{array}{ll}
  H^g_m(s_0)\simeq H^g_m(\gamma, H^{(0)}(\delta))\,,& g\geq m\geq 0;
  \\[3mm]
  H^g_m(s_0|D)\simeq H^g_m(\gamma, H^{(0)}(\delta|D))\,,& g\geq m\geq 0;
  \\[3mm]
  H^g_{g+1}(s_0|D)\simeq H^{(1)}_{g+1}(\delta|D)\,, & g\geq -1\,. \\
\end{array}
$$
From the viewpoint of physics, the most notable among these groups are the following:
\begin{itemize}
    \item $H^0_0(s_0)$ the group of physical observables with values
    in local functions;
    \item $H^0_0(s_0|D)$ the group of physical observables with
    values in local functionals;
    \item $H^{-1}_0(s_0|D)$ the group of characteristics;
    \item $H^{0}_1(s_0|D)$ the group of rigid symmetries;
    \item $H^{1}_2(s_0|D)$ the group of Lagrange structures;
    \item $H^{2}_3(s_0|D)$ the group of potential obstructions to
    integrability of the Lagrange structures.
\end{itemize}

In the next section we study all these groups  more closely.

\section{The local BRST cohomology of ODEs}

\subsection{The group $H(\delta)$} The algebraic concept of \textit{filtration} \cite{Mac}
considerably facilitates (or even makes possible) the computation of
(co)homology groups. In our geometric setting, it comes from the
natural filtration of the underlying jet space. Namely, let us
arrange  the variables coordinatizing  the vertical part of the
infinite jet space $J^\infty V$ in the following increasing sequence
of finite sets:
$$
    \begin{array}{l}
      V_0=\{x^i, \bar\eta_i, c^\alpha, \bar\xi{}^a,\eta_a,\bar\lambda_\alpha\}\,,
      \\[3mm]
      V_s=V_{s-1}\cup \Big\{ \stackrel{_{(s)}}{x}{\!\!}^i, \stackrel{_{(s-1)}}{\bar x}{\!\!}_i, \stackrel{_{(s-1)}}{\eta}{\!\!}^i,
     \stackrel{_{(s)}}{\bar\eta}{\!\!}_i,
    \stackrel{_{(s-1)}}{\bar\eta}{\!\!}^a, \stackrel{_{(s)}}{\eta}{\!\!}_a,    \stackrel{_{(s-1)}}{
    \lambda}{\!\!}^\alpha, \stackrel{_{(s)}}{
    \bar\lambda}{\!\!}_\alpha, \stackrel{_{(s-1)}}{
    \xi}{\!\!\!\!}_a,\stackrel{_{(s)}}{\bar
    \xi}{\!\!}^a, \stackrel{_{(s)}}{
    c}{\!\!}^\alpha,  \stackrel{_{(s-1)}}{\bar
    c}{\!\!\!\!}_\alpha\Big\}\,,\quad s\in \mathbb{N}\,.
    \end{array}
$$
Associated to this sequence  is the ascending filtration of the
space of local functions
\begin{equation}\label{F-FILTR}
F_0\subset F_1\subset F_2\subset \cdots\subset F_\infty=F\,,
\qquad F_s=C^\infty(V_s)\,.
\end{equation}
The filtration  is chosen so as to be compatible with the action
of the Koszul-Tate differential, that is, $\delta F_s\subset F_s$.
The last property is easily  seen from the following explicit
expressions\footnote{Notice that the most natural filtration of
$F$ with $F_k=C^\infty(J^k V)$ is not respected by $\delta$.}:
$$
\begin{array}{ll}
\delta
\stackrel{_{(k)}}{\eta}{\!\!}^i=\stackrel{_{(k+1)}}{x}{\!\!}^i+D^k(V^i+\lambda^\alpha
R_\alpha^i)\,,& \displaystyle \delta
\stackrel{_{(k)}}{\xi}{\!\!}_a=D^{k}\left(-
\eta_{b}D^{b}_{a}-\lambda^{\alpha}A_{\alpha
a}^{b}\eta_{b}+\eta^{i}\frac{\partial T_{a}}{\partial
x^{i}}\right)-\stackrel{_{(k+1)}}{\eta}{\!\!\!\!}_{a}\,.\\[5mm]
\delta \stackrel{_{(k)}}{\eta}{\!\!}_a=D^k T_{a}\,,\quad\delta
\stackrel{_{(k)}}{\bar{\lambda}}{\!\!}_\alpha=-D^k(\bar{\eta}_i
R_\alpha^i)\,,&\displaystyle \delta \stackrel{_{(k)}}{\bar
x}{\!\!}_i=-D^k\left(\frac{\partial V^j}{\partial x^i}\bar
\eta_j+\lambda^{\alpha}\frac{\partial R_{\alpha}^{j}}{\partial
x^{i}}\bar{\eta}_{j}+\bar{\eta}^{a}\frac{\partial T_{a}}{\partial
x^{i}}\right)+\stackrel{_{(k+1)}}{\bar\eta}{\!\!\!\!}_i\,,
\end{array}
$$
$$
\delta \stackrel{_{(k)}}{\bar{c}}{\!\!}_\alpha
=\stackrel{_{(k+1)}}{\bar{\lambda}}{\!\!\!\!}_{\alpha}+D^{k}\left(R_{\alpha}^{i}\bar{x}_{i}+
\bar\lambda{}_\beta
E^\beta_\alpha+\lambda^{\beta}B_{\beta\alpha}^{\gamma}\bar{\lambda}_{\gamma}+\lambda^{\beta}C_{\alpha\beta}
^{ia}\bar{\eta}_{i}\eta_{a}-F_{\alpha}^{ia}\bar{\eta}_{i}\eta_{a}-A_{\alpha
a}^{b}\bar{\eta}^{a}\eta_{b}-\frac{\partial R_{\alpha}^{i}}{\partial
x^{j}}\bar{\eta}_{i}\eta^{j}\right).
$$
(The other variables of $V_s$ are annihilated by $\delta$.) Here
we introduced the Cartan vector field on $J^\infty V$,
\begin{equation}\label{D}
D=\sum_{s=0}^\infty
\stackrel{_{(s+1)}}{\Phi}{\!\!}^A\frac{\partial}{\partial
\stackrel{_{(s)}}{\Phi}{\!\!}^A}\,,
\end{equation}
which is nothing else but the jet counterpart of the operator of
time derivative. It should be noted that the action of the
Koszul-Tate differential defines (and is defined by) the boundary
condition for the classical BRST differential \cite{KLS2}. So, no
other terms than those written explicitly down in (\ref{BOUND})
are needed to find the action of $\delta$ in $V_s$.

The filtration (\ref{F-FILTR}) is exhaustive in the sense that
each local function belongs to some $F_s$ for $s$ large enough.
The natural inclusions $i_{ss'}: F_s\rightarrow F_{s'}$ for $s\leq
s'$ induce the homomorphisms $i^\ast_{ss'}: H( F_s)\rightarrow
H(F_{s'})$ of the homology groups associated with the direct
system of complexes $\{F_s, i_{ss'}\}$ indexed by $\mathbb{N}$. As
the homology functor commutes with direct limits\footnote{ Even
$\displaystyle \lim_{\rightarrow} F_s$ is just a union, $\{ H(
F_s), i^\ast_{ss'}\}$ is generally a nontrivial direct system as
the homology functor does not preserve monomorphisms.}, we can
define the $\delta$-homology groups of the complex $F$ by setting
$\displaystyle H (\delta)=\lim_{\rightarrow}H( F_s) $. By
definition of the direct limit any element of $H (\delta)$ is
represented by a cycle that belongs to at least one space $F_s$.

Since each complex $F_s$ consists of smooth functions living on a
\textit{finite-dimensional} graded superdomain, we can freely
apply to them all the usual differential-geometric constructions
like the inverse function theorem. In particular, consider the
change of variables $V_s$ whereby
$$
\begin{array}{lll}
\stackrel{_{(k+1)}}{x}{\!\!}^i \quad \mapsto\quad
\stackrel{_{k+1}}{x}{\!\!}^i=\delta
\stackrel{_{(k)}}{\eta}{\!\!}^i \,,\quad&
\stackrel{_{(k+1)}}{\eta}{\!\!}_a\quad \mapsto\quad
\stackrel{_{k+1}}{\eta}{\!\!}_a=\delta
\stackrel{_{(k)}}{\xi}{\!\!}_a\,,&\\[3mm]
\stackrel{_{(k+1)}}{\bar\eta}{\!\!}_i\quad \mapsto\quad
\stackrel{_{k+1}}{\bar\eta}{\!\!}_i=\delta \stackrel{_{(k)}}{\bar
x}{\!\!}_i\,,\quad &
 \stackrel{_{(k+1)}}{\bar\lambda}{\!\!}_\alpha \quad\mapsto \quad   \stackrel{_{k+1}}{\bar\lambda}{\!\!}_\alpha=\delta
\stackrel{_{(k)}}{c}{\!\!}_\alpha\,, \quad& k=0,1,\ldots, s-1\,,
\end{array}
$$
and all the other variables of $V_s$ remain the same. It is easy
to see that this change of coordinate variables is nondegenerate
and brings the Koszul-Tate differential to the form
$$
 \delta|_{F_s} =\sum_{k=0}^{s-1} \left(\stackrel{_{k+1}}{x}{\!\!}^i\frac{\partial}{\partial \stackrel{_{(k)}}{\eta}{\!}^i}
     +\stackrel{_{k+1}}{\bar\eta}{\!\!\!}_i\frac{\partial}{\partial
     \stackrel{_{(k)}}{\bar x}{\!}_i}+\stackrel{_{k+1}}{\bar\lambda}{\!\!\!}_\alpha\frac{\partial}{\partial
     \stackrel{_{(k)}}{\bar c}{\!}_\alpha}-\stackrel{_{k+1}}{\eta}{\!\!\!}_{a}\frac{\partial}{\partial
\stackrel{_{(k)}}{\xi}{\!}_{a}}\right)-
\stackrel{}{\bar{\eta}}{\!\!}_i
R_\alpha^i(\stackrel{}{x})\frac{\partial}{\partial
{\stackrel{}{\bar{\lambda}}_{\alpha}}}+T_{a}(\stackrel{}{x})\frac{\partial}{\partial
{\stackrel{}{\eta}_{a}}}\,.
$$
(For $s=0$ the first sum is absent.) Define the sequence of  sets
$$
V_0^0=V_0\,,\qquad V^0_s=V^0_{s-1} \cup
\Big\{\stackrel{_{(s-1)}}{\lambda}{\!\!\!}^\alpha,
\stackrel{_{(s-1)}}{\bar\eta}{\!\!\!}^a,
\stackrel{_{(s)}}{c}{\!\!}^{\alpha},\stackrel{_{(s)}}{\bar{\xi}}{\!\!}^{a}\Big\}\,,\quad
s\in \mathbb{N}\,.
$$
The complex $F_s$ splits into the direct sum $F_s^0\oplus F'_s$ of
two subcomplexes, where the elements of $F^0_s$ are the smooth
functions of the variables $V_s^0$. Writing $\pi: F_s\rightarrow
F_s^0$ for the natural projection that takes all the variables
from $V_s\backslash V_s^0$ to zero,  we can define the
complementary subspace as $F_s'=(1-\pi)F_s$. The invariance of
$F^0_s$ and $F'_s$ under the action of $\delta$ is obvious.

Now we claim that the complex $F'_s$ is acyclic. Indeed,  consider
the operator
$$
    \sigma=\sum_{k=0}^{s-1}
    \left(\stackrel{_{(k)}}{\eta}{\!}^i\frac{\partial}{\partial \stackrel{_{k+1}}{x}{\!\!}^{i}}+\stackrel {_{(k)}}{\bar x}{\!\!}_{i}
    \frac{\partial}{\partial \stackrel{_{k+1}}{\bar \eta}{\!\!\!}_i}+\stackrel{_{(k)}}{\bar c}{\!\!}_\alpha\frac{\partial}{\partial
    \stackrel{_{k+1}}{\bar\lambda}{\!\!}_\alpha} - \stackrel{_{(k)}}{\xi}{\!\!}_a\frac{\partial}{\partial
    \stackrel{_{k+1}}{\eta}{\!\!\!}_a}\right)\,,
$$
which maps  $F'_s$ into itself  and squares to zero. The
anti-commutator of $\sigma$ and $\delta$ is given by
$$
    N=\sum_{k=0}^{s-1} \left(\stackrel{_{k+1}}{x}{\!\!}^i\frac{\partial}{\partial \stackrel{_{k+1}}{x}{\!\!}^{i}}+
    \stackrel {_{k+1}}{\bar \eta}{\!\!\!}_{i}
    \frac{\partial}{\partial \stackrel{_{k+1}}{\bar \eta}{\!\!\!}_i}+\stackrel{_{k}}{\eta}{\!}^i\frac{\partial}{\partial \stackrel{_{k}}{\eta}
    {\!}^{i}}+\stackrel {_{k}}{\bar x}{\!\!}_{i}
    \frac{\partial}{\partial \stackrel{_{k}}{\bar x}{\!\!}_i}+\stackrel {_{k}}{\bar c}{\!}_{\alpha}
    \frac{\partial}{\partial \stackrel{_{k}}{\bar c}{\!}_\alpha}+\stackrel {_{k+1}}{\bar \lambda}{\!\!}_{\alpha}
    \frac{\partial}{\partial \stackrel{_{k+1}}{\bar \lambda}{\!\!}_\alpha}+\stackrel {_{k}}{\xi}{\!\!}_{a}
    \frac{\partial}{\partial \stackrel{_{k}}{\xi}{\!\!}_a}+\stackrel {_{k+1}}{\eta}{\!\!\!}_{a}
    \frac{\partial}{\partial
    \stackrel{_{k+1}}{\eta}{\!\!\!}_a}\right).
$$
Since $N$ is obviously invertible in $F'_s$, the composition
$h=N^{-1}\sigma$ gives a contracting homotopy for $\delta|_{F'}$
and acyclicity of $F'_s$ follows.

The restriction of the Koszul-Tate differential to $F_s^0$ is
given by the operator
$$
\delta|_{F_s^0}=T_{a}\frac{\partial}{\partial
\stackrel{_{}}{\eta_{a}}}-\bar{\eta}_i
R_\alpha^i\frac{\partial}{\partial
\stackrel{_{}}{\bar{\lambda}_{\alpha}}}\,,
$$
which homology can be described as follows. Due to the
irreducibility conditions (\ref{IRRED}) for the gauge symmetries
and constraints, a function $a\in F^0_s$ is a $\delta$-cycle iff
it is independent of $\eta_a$'s and $\bar\lambda$'s. Let
$\widetilde{F}_s^0$ denoted the algebra of smooth functions of the
variables $\widetilde{V}{}^0_s=V^0_s\backslash
\{\bar\lambda_\alpha,\eta_a\}$ and let $I_s$ be an ideal in
$\widetilde{F}{}^0_s$ generated by the functions $T_a(x)$ and
$\bar\eta_iR^i_\alpha(x)$.  It is clear that $I_s=\delta F_s^0\cap
\widetilde{F}^0_s$, and hence
\begin{equation}\label{H(delta)}
H(F_s)\simeq H(F_s^0) \simeq\widetilde{F}^0_s/I_s\,.
\end{equation}
One can also give the group $H( F_s)$  a geometrical
interpretation. The ideal $I_s$, being regular, defines a smooth
submanifold $M_s$ in the superdomain with coordinates
$\widetilde{V}{}^0_s$. The ``points'' of $M_s$ are solutions to
the equations
$$
T_a(x)=0\,,\qquad \bar\eta_iR^i_\alpha(x)=0\,.
$$
Then (\ref{H(delta)}) says that the group $H( F_s)$ is isomorphic
to the space of smooth functions on $M_s$.

Notice that all the elements of $H( F^0_s)$ have resolution degree
zero. This is in agreement with the general property of the
Koszul-Tate differential of being acyclic in positive resolution
degree.

\subsection{The group $H(\gamma, H(\delta) )$} Having studied the
$\delta$-homology, we can now turn to the cohomology associated with
the longitudinal differential. To do this requires an explicit
expression for the action of $\gamma$ on local functions. Unlike the
Koszul-Tate differential, the Hamiltonian action of the boundary
terms (\ref{BOUND}) by themselves do not specify the whole $\gamma$.
In this case we need to know the classical BRST charge up to the
second order in resolution degree. The missing terms of resolution
degree $2$ can  easily be found from the classical master equation
by means of the homological perturbation theory. Without going into
detail we simply present the function that should be added to
(\ref{BOUND}) to have $\Omega_1$ specified up to the second order in
resolution degree. It reads
\begin{equation}\label{ADD}
\bar{c}_{\alpha}\Psi^{\alpha}+\xi_{a}\Theta^{a}+\bar{\lambda}_{\alpha}\tau\Psi^{\alpha}+
\eta_{a}\tau\Theta^{a}\,,
\end{equation}
where
$$
\Psi^{\alpha}=\frac{1}{2}c^{\gamma}c^{\beta}B_{\beta\gamma}^{\alpha},\qquad
\Theta^{a}=\frac{1}{2}c^{\beta}c^{\gamma}C_{\beta\gamma}^{ai}\bar{\eta}_{i}+
c^{\beta}\bar{\xi}^{b}A_{\beta b}^{a}\,,
$$
and we introduced the operator
\begin{equation}\label{TAU}
\tau =\eta^i\frac{\partial }{\partial x^i}+\bar x_i\frac{\partial
}{\partial \bar \eta_i}\,.
\end{equation}
Notice that expression (\ref{ADD}) (and hence, the second order BRST
charge)  involves only the structure functions of the involutivity
conditions (\ref{INV}). Of course, higher orders in resolution
degree, if any, involve new structure functions coming from the
iterated commutators of $V$,  $R$'s, and $T$'s. Important though
these higher structure functions are for the definition of the
classical BRST complex, they  do not contribute to (the computation
of) the classical BRST cohomology.

Summing (\ref{BOUND}) and (\ref{ADD}) and extracting the
zero-resolution-degree part in the classical BRST differential
(\ref{BRST-DIFFERENTIAL}), we find
\begin{equation}\label{LD}\begin{array}{l}
    \displaystyle\gamma \stackrel{_{(k)}}{x}{\!\!}^{i}=D^{k}\left(c^{\alpha}R_{\alpha}^{i}\right)\,,
    \qquad \displaystyle\gamma\stackrel{_{(k)}}{\lambda}{\!\!}^{\alpha}=D^k\left(c^{\beta}E_{\beta}^{\alpha}+
    c^{\gamma}\lambda^{\beta}B^{\alpha}_{\beta\gamma}\right)-\stackrel{_{(k+1)}}{c}{\!\!}^{\alpha}\,,\\[4mm]
    \displaystyle\gamma\stackrel{_{(k)}}{c}{\!\!}^{\alpha}=D^{k}\Psi^{\alpha}\,,\qquad\qquad
    \displaystyle\gamma\stackrel{_{(k)}}{\bar{\eta}}{\!\!}_{i}=
    D^k\left(\bar{\xi}^a\frac{\partial T_a}{\partial x^i}-c^{\alpha}\frac{\partial R_{\alpha}^{j}}{\partial x^{i}}\bar{\eta}_{j}\right)\,,\qquad
     \displaystyle\gamma\stackrel{_{(k)}}{\bar{\xi}}{\!\!}^{a}=-D^{k}\Theta^{a}\,,\\[4mm]
    \displaystyle\gamma\stackrel{_{(k)}}{\bar{\eta}}{\!\!}^{a}=D^{k}\left(c^{\alpha}\lambda^{\beta}C_{\alpha\beta}^{ia}\bar{\eta}_{i}
    -c^{\alpha}F_{\alpha}^{ia}\bar{\eta}_{i}
    -c^{\alpha}A_{\alpha b}^{a}\bar{\eta}^{b}-\bar{\xi}^{b}D_{b}^{a}-\bar{\xi}^b\lambda^{\alpha}A_{\alpha
    b}^{a}\right)+\stackrel{_{(k+1)}}{\bar{\xi}}{\!\!\!}^{a}\,,
    \\[4mm]\displaystyle
    \displaystyle\gamma \stackrel{_{(k)}}{\bar{x}}{\!\!}_{i}=
    -D^{k}\left[c^{\alpha}\frac{\partial}{\partial x^i}\left(R_{\alpha}^{j}\bar{x}_{j}+
    \bar\lambda{}_\beta E^\beta_\alpha+\lambda^{\beta}B_{\beta\alpha}^{\gamma}\bar{\lambda}_{\gamma}+\lambda^{\beta}C_{\alpha\beta}
    ^{aj}\bar{\eta}_{j}\eta_{a}-F_{\alpha}^{aj}\bar{\eta}_{j}\eta_{a}-A_{\alpha
    a}^{b}\bar{\eta}^{a}\eta_{b}\right.\right.\\[4mm]\displaystyle\left.\left.
    -\frac{\partial R_{\alpha}^{m}}{\partial
    x^{j}}\bar{\eta}_{m}\eta^{j}\right)
    -\bar{\xi}^{a}\frac{\partial}{\partial x^i}\left( \eta_{b}D^{b}_{a}+\lambda^{\alpha}A_{\alpha
    a}^{b}\eta_{b}-\eta^{j}\frac{\partial T_{a}}{\partial x^{j}}\right)\right]\,,\\[4mm]
    \gamma\stackrel{_{(k)}}{\bar{\lambda}}{\!\!}_{\alpha}=
    D^k\left(\bar\xi^a A_{\alpha a}^b\eta_b - c^{\gamma}\bar{\lambda}_{\beta}
    B^{\beta}_{\alpha\gamma}-
    c^\gamma C_{\gamma\alpha}^{ai}\bar\eta_i\eta_a\right)\,,\quad  \displaystyle \gamma\stackrel{_{(k)}}{\xi}{\!\!}_{a}=D^{k}
    \Big(\xi_b c^\beta A_{\beta a}^b+
    \eta_b c^\beta \eta^i\frac{\partial A_{\beta a}^b}{\partial x^i}\Big)\,,\\[4mm]
    \displaystyle\gamma\stackrel{_{(k)}}{\bar{c}}{\!\!}_{\alpha}=D^{k}\left(\bar{c}_\beta c^\gamma B_{\alpha\gamma}^\beta-
    \xi_a c^\gamma C^{ai}_{\gamma\alpha}\bar\eta_i+A_{\alpha b}^a\bar{\xi}^b \xi_a-
    \bar{\lambda}_\beta c^\gamma\eta^i\frac{\partial B_{\alpha\gamma}^{\beta}}{\partial x^i}+\eta_a c^{\gamma} \eta^j
    \frac{\partial C_{\gamma\alpha}^{ai}}{\partial x^j}\bar{\eta}_i\right.\\[4mm]\left.+\eta_a c^{\gamma} C_{\gamma\alpha}^{ai}\bar{x}_i-
    \displaystyle \eta_a \eta^i\frac{\partial A_{\alpha b}^a}{\partial x^i} \bar{\xi}^b\right)\,,\\[4mm]
    \displaystyle\gamma\stackrel{_{(k)}}{\eta}{\!\!}^{i}=D^k\left(-c^\alpha\lambda^\beta C_{\alpha\beta}^{ai}\eta_a+
    c^\alpha F_{\alpha}^{ai}\eta_a+c^\alpha \frac{\partial R_\alpha^i}{\partial
    x^j}\eta^j\right)\,,
    \qquad \gamma \stackrel{_{(k)}}{\eta}{\!\!}_{a}=D^k\left(c^\alpha
A_{\alpha a}^b\eta_b\right)\,.
\end{array}\end{equation}
As is seen, $\gamma$ respects the filtration (\ref{F-FILTR}) in
the sense that $ \gamma F_s\subset F_s$ and we can set
$\displaystyle H(\gamma, H(\delta))=\lim_{\rightarrow}H(\gamma,
H(F_s))$.

Let us now show that the complex $(\gamma, H(\delta))$ is
homotopic to its subcomplex $(\gamma, H(F_0^0))$ so that
$H(\gamma, H(\delta))\simeq H(\gamma, H(F^0_0))$. To this end,
introduce the operator
$$
\sigma_s=\stackrel{_{(s-1)}}{\bar\eta}{\!\!\!}^a\frac{\partial
}{\partial
\stackrel{_{(s)}}{\bar\xi}{\!\!}^a}-\stackrel{_{(s-1)}}{\lambda}{\!\!\!}^\alpha\frac{\partial
}{\partial \stackrel{_{(s)}}{c}{\!\!}^\alpha}\,,
$$
which maps $F_s$ into itself. It is easy to see that $\sigma_s$
anti-commutes with $\delta$, inducing a well-defined operator in
$H(F_s)$. Anti-commuting $\sigma_s$ with $\gamma$, we get
$$
    N_s=[\sigma_s,{\gamma}]=\stackrel{_{(s)}}{c}{\!\!}^{\alpha}\frac{\partial
    }{\partial \stackrel{_{(s)}}{c}{\!\!}^{\alpha}}+\stackrel{_{(s-1)}}{\lambda}{\!\!\!}^{\alpha}\frac{\partial
    }{\partial \stackrel{_{(s-1)}}{\lambda}{\!\!\!}^{\alpha}}+\stackrel{_{(s)}}{\bar{\xi}}{\!\!}^{a}\frac{\partial
    }{\partial \stackrel{_{(s)}}{\bar{\xi}}{\!\!}^{a}}+\stackrel{_{(s-1)}}{\bar{\eta}}{\!\!\!}^{a}\frac{\partial
    }{\partial \stackrel{_{(s-1)}}{\bar{\eta}}{\!\!\!}^{a}}\,.
$$
The nontrivial ${\gamma}$-cocycles are bound to center in the
kernel of the operator $N_s$. Since $\mathrm{Ker}\,
N_s={F}_{s-1}$, we infer that any ${\gamma}$-cocycle from $H(F_s)$
is cohomlogous to one from $H({F}_{s-1})$ and,  by induction,
$H(\gamma, H(F_s))\simeq H(\gamma, H(F_0))$. It remains to note
that according to (\ref{H(delta)}) $$H(F_0)\simeq H(F^0_0)\simeq
\widetilde{F}{}^0_0/I_0\,.$$

By definition, $\widetilde{F}_0^0$ is the algebra constituted by
the smooth functions of the variables
$\widetilde{V}{}^0_0=\{x^i,\bar\eta_i,c^\alpha,\bar\xi{}^a\}$ and
the ideal $I_0\subset \widetilde{F}_0^0$ is generated by the
functions $T_a$ and $\bar\eta_iR^i_\alpha$. It follows from
(\ref{LD}) that both $\widetilde{F}_0^0$ and $I_0$ are invariant
under the action of the longitudinal differential and we can study
the $\gamma$-cohomology directly in the quotient space
$\widetilde{F}{}^0_0/I_0$. Letting
$\gamma_0=\gamma|_{\widetilde{F}{}^0_0}$, we find
$$
\gamma_0=c^{\alpha}R_{\alpha}^{i}\frac{\partial }{\partial
x^i}+\left(\bar{\xi}^a\frac{\partial T_a}{\partial
x^i}-c^{\alpha}\bar{\eta}_{j}\frac{\partial
R_{\alpha}^{j}}{\partial x^{i}}\right)\frac{\partial }{\partial
\bar\eta_i}+\frac{1}{2}c^{\gamma}c^{\beta}B_{\beta\gamma}^{\alpha}\frac{\partial
}{\partial
c^\alpha}+\left(\frac{1}{2}c^{\beta}c^{\gamma}C_{\beta\gamma}^{ia}\bar{\eta}_{i}+
c^{\beta}\bar{\xi}^{b}A_{\beta b}^{a}\right)\frac{\partial
}{\partial \bar\xi^a}\,.
$$
The inclusion $\gamma_0 I_0\subset I_0$ follows immediately from
the involutivity conditions (\ref{INV}). By definition, a class
$a+I_0$  is a $\gamma_0$-cocycle iff $\gamma_0 a\in I_0$; it is a
$\gamma_0$-coboundary iff $a=\gamma_0 b+c$ for some $b\in
\widetilde{F}{}^0_0$ and $c\in I_0$. This leads us to  the
identification
\begin{equation}\label{ISO-G}
    H(\gamma, H(\delta))\simeq \frac{\gamma^{-1}_0 I_0}{\mathrm{Im}\,\gamma_0\cup
    I_0}\,.
\end{equation}

Consider, for example, the $\gamma$-cohomology in ghost number
zero. As the ghost numbers of the variables $\bar\eta_i$,
$c^\alpha$, and $\bar\xi{}^a$ are strictly positive,  the
representative cocycles are given by the functions of $x$'s
considered modulo constraints $T_a$. A function $a(x)$ defines a
$\gamma$-cocycle if
\begin{equation}\label{gadb}
    c^{\alpha}R_{\alpha}^{i}\frac{\partial a}{\partial
    x^{i}}=c^{\alpha}U_{\alpha}^{a}T_{a}
\end{equation}
for some smooth functions  $U^a_\alpha(x)$. In other words, the
cocycle $a(x)$ is to be annihilated by the gauge distribution
$R=\{R_\alpha\}$ on the constraint surface $\Sigma$, and two such
cocycles are equivalent iff their difference vanishes on $\Sigma$.
This is exactly the definition of the $t$-local physical
observables we have discussed in Sec. 2. The physical role of the
other groups, $H^g(\gamma, H^0(\delta))$ with $g>0$, is not well
understood.

\subsection{The group $H(\delta|D)$}
Now we proceed to the study of the relative $\delta$-homology. As
before, our main computational tool is the concept of filtration.
A reliant filtration here is, of course, that induced by
(\ref{F-FILTR}). The inclusions $i_{ss'}: F_s\rightarrow F_{s'}$
underlying the filtration (\ref{F-FILTR}) pass trough the quotient
$F/DF$ giving rise to the direct system of complexes $\{F_s/DF_s,
j_{ss'}\}$ with $j_{ss'}$ induced by $i_{ss'}$. This allows us to
define the $\delta$ modulo $D$ homology groups as the direct limit
$\displaystyle H(\delta|D)=\lim_{\rightarrow} H(\delta,
F_s/DF_{s-1})$. Let $D_s: F_{s}\rightarrow F_{s+1}$ denote the
restriction of the operator (\ref{D}) onto $F_s$. The operator
$D_s$ being a chain transformation, we have  two short exact
sequences of complexes
\begin{equation}\label{AB}
\begin{array}{ll}
A:&\xymatrix{0 \ar[r]&{\mathrm{Ker}D_s}\ar[r]^-{i_1}&{F_s
}\ar[r]^-{p_1} & \mathrm{Im} D_s \ar[r]&0}\,,\\[3mm]
B:& \xymatrix{0 \ar[r]&{\mathrm{Im}D_s}\ar[r]^-{i_2}&{F_{s+1}
}\ar[r]^-{p_2} & \mathrm{Coker} D_s \ar[r]&0}\,.
\end{array}
\end{equation}
With this notation the group $H(\delta|D)$ is given by the direct
limit $\displaystyle \lim_{\rightarrow}H(\mathrm{Coker}D_s)$.

There is a standard algebraic construction \cite[p.46]{Mac} that
allows one to fit the induced map $D_s^\ast: H( F_s)\rightarrow H(
F_{s+1})$ on homology into an exact sequence that relates the
groups $H(F_s)$, $H(F_{s+1})$, and $H(\mathrm{Coker}D_s)$. The
construction goes as follows. First, one defines the
\textit{mapping cone} of the chain transformation $D$ to be the
chain complex $\mathrm{Con} D_s=F_{s}\oplus F_{s+1}$ with
differential
$$
\delta (a,b)=(-\delta a, D_sa +\delta b)\,,\qquad
\deg\,(a,b)=\mathrm{deg}\, b=\mathrm{deg}\, a+1\,.
$$
Since $\mathrm{Ker}D=\mathbb{R}$ and $\mathrm{Im}\,\delta\cap
\mathrm{Ker}\, D=0$, we conclude that
\begin{equation}\label{DS}
H(\mathrm{Con}D_s)\simeq H(\mathrm{Coker} D_s)\oplus \mathbb{R}\,,
\end{equation}
where the second summand is generated by the $1$-cycle $(1,0)$.
The natural injection $i: F_{s+1}\rightarrow \mathrm{Con} D_s$ is
a cochain transformation. The projection $p: \mathrm{Con}
D_s\rightarrow \bar{F}_s$ with $p(a,b)=a$ is also a chain
transformation, if by $\bar{F}_s$ we mean the complex $F_s$ with
the dimensions all lowered by 1 and differential $-\delta$. Thus
we arrive at the short exact sequence of complexes
\begin{equation*}
\xymatrix{0 \ar[r]&{F_{s+1}}\ar[r]^-{i}&{\mathrm{Con} D_s
}\ar[r]^-{p} & \bar{F}_s \ar[r]&0}\,.
\end{equation*}
It is clear that $H(F_s)\simeq H(\bar F_s)$ as vector spaces. The
short exact sequence above gives rise to the triangle diagram
\begin{equation}\label{D1}
\xymatrix{H(F_{s})\ar[rr]^{D_s^\ast}& & H(F_{s+1}) \ar[dl]^{i_\ast}\\
&H(\mathrm{Con} D_s)\ar[ul]^{p_\ast}& }
\end{equation}
with exact vertices and $D_s^\ast$ playing  the role of the
connecting homomorphism. Since we are dealing with complexes of
vector spaces, the triangle diagram implies the existence of an
isomorphism
$$
H(\mathrm{Con}D_s)\simeq\mathrm{Ker}D_s^\ast\oplus
\mathrm{Coker}D_s^\ast
$$
and it remains to compute the kernel and cokernel of the operator
$D_s^\ast$.

\vspace{3mm} \noindent
 \textit{Remark}. {There is also another
triangle diagram canonically associated to the mapping cone
\begin{equation}\label{D2}
\xymatrix{H(\mathrm{Coker} D_s)\ar[rr]^-{\partial}& & H(\mathrm{Ker}D_s) \ar[dl]^-{j_\ast}\\
&H(\mathrm{Con}D_s)\ar[ul]^-{k_\ast}& }
\end{equation}
This diagram is exact although the sequence  of complexes
$$\xymatrix{0\ar[r]& {\mathrm{Ker}D_s} \ar[r]^-{j}&
{\mathrm{Con}D_s}\ar[r]^-{k}& {\mathrm{Coker} D_s}\ar[r]& 0}$$ is
not; here $ja=(i_1a,0)$ for $a\in \mathrm{Ker}D_s$, $k(a,b)=p_2b$,
and  $\partial=\partial_A\partial_B$ is the composition of
connecting homomorphisms for (\ref{AB}). Since $\mathrm{Ker}
D_s=\mathbb{R}= H(\mathrm{Ker}D_s)$, the group
$H(\mathrm{Con}D_s)$ splits as in (\ref{DS}).} \vspace{3mm}

Let us start with the space $\mathrm{Ker} D^\ast_s$. We know that
$H(F_s)\simeq \widetilde{F}{}^0_s/I_s$, so that each class of
$\delta$-homology is represented by some function $f\in
\widetilde{F}{}^0_s$. Introducing the collective notation for the
coordinates
$$z_0^A=(c^\alpha,\bar\xi{}^a)\,,\qquad
z^A_s=\Big(\stackrel{_{(s-1)}}{\lambda}{\!\!}^\alpha,\stackrel{_{(s-1)}}{\bar\eta}{\!\!}^a,
\stackrel{_{(s)}}{c}{\!\!}^{\alpha},\stackrel{_{(s)}}{\bar{\xi}}{\!\!}^{a}\Big)\,,\qquad
s\in \mathbb{N}\,,$$ we can write the action of the operator $D_s$
on $f\in \widetilde{F}^0_s$ as
\begin{equation}\label{Df}
    D_sf= D_s'f+ Lf+\delta(\tau f)\,,
\end{equation}
where
$$
   D_s'=\sum_{k=0}^{s}z^A_{k+1}\frac{\partial }{\partial
z^A_k}\,,\qquad
 L=-V^{i}\frac{\partial }{\partial x^{i}}-
    \lambda^{\alpha}R_{\alpha}^{i}\frac{\partial }{\partial x^{i}}+
    \bar{\eta}_{i}\frac{\partial V^{i}}{\partial x^{j}}\frac{\partial }{\partial
    \bar{\eta}_{j}}+
    \lambda^{\alpha}\bar{\eta}_{i}\frac{\partial R^{i}_{\alpha}}{\partial x^{j}}\frac{\partial }{\partial
    \bar{\eta}_{j}}+
    \bar\eta{}^a\frac{\partial T_a}
    {\partial x^i}\frac{\partial }{\partial \bar\eta_i}\,,
$$
and the operator $\tau$ is defined by (\ref{TAU}). Notice that the
operator $D'_s+L$ maps the algebra $\widetilde{F}{}^0_{s}$ into
$\widetilde{F}{}^0_{s+1}$ and the ideal $I_s$ into $I_{s+1}$;
hence, its action descends  to the $\delta$-homology. Then the
condition $[f]\in \mathrm{Ker}D_s^\ast$ implies that
\begin{equation}\label{kerD}
D'_sf+Lf=g
\end{equation}
for some $g\in I_{s+1}$. Applying now the operator
$$
\varrho_s=z^A_{s}\frac{\partial }{\partial z^A_{s+1}}
$$
to both sides of equation (\ref{kerD}), we get
$$
z^A_{s}\frac{\partial f}{\partial z^A_{s}}=\varrho_sg
$$
or, what is the same,
$$
t\frac{d}{dt} f(tz^A_s)=(\varrho_sg)(tz^A_s)\,.
$$
This yields
$$
f(z^A_s)-f(0)=\int_{0}^1\frac{dt}{t}(\varrho_sg)(tz^A_s)\in
I_{s}\,.
$$
The function $\varrho_sg$ being proportional to $z^A_s$, the
integral is well-defined and  $f$ appears to be  homologous to
some $f_0\in \widetilde{F}^0_{s-1}$. Proceeding in this way we see
that $f$ is cohomologous to a function with no dependence of
$z$'s, i.e., to a function of the variables $x^i$ and
$\bar\eta_i$. Let $\widetilde{F}^0_{-1}$ denote the space of such
functions. For any $f\in \widetilde{F}{}^0_{-1}$ equation
(\ref{kerD}) reduces to
\begin{equation}\label{PrCon}
Lf=g\,.
\end{equation}
Since  $\lambda$'s and $\bar\eta^a$'s enter the l.h.s. of
(\ref{PrCon}) at most linearly, we can always take $g=\delta h$
with
$$
h=\bar{\lambda}_{\alpha}(h^{\alpha}+\lambda^{\beta}h^{\alpha}_{\beta}+\bar\eta^ah_a^\alpha)+
    \eta_{a}(h^{a}+\lambda^{\beta}h^{a}_{\beta}+\bar\eta^bh_b^a)
$$
and $h$'s being functions of $x^i$ and $\bar\eta_i$. Denoting by
$L_0$ the restriction of the operator $L$ onto
$\widetilde{F}{}^0_{-1}$ we can summarize our consideration by the
following compact formula
\begin{equation}\label{KerD}
    \mathrm{Ker}\,D^\ast_s\simeq L_0^{-1}I_1\,.
\end{equation}
The structure of the last isomorphism  is easily unfolded by
reading formula (\ref{Df}) from right to left. It just says that
if $f\in \widetilde{F}_{-1}^0$ satisfies (\ref{PrCon}), that is,
belongs to $L_0^{-1}I_1$, then the function $\tau f+h$ is a
$\delta$ modulo $D$ cycle. Since
$\mathrm{Ker}\,\tau|_{\widetilde{F}^0_{-1}}=\mathbb{R}$, the
relative  cycle $\tau f+h$ is nontrivial whenever the function $f$
is nonconstant. It is the space of constant functions that
corresponds to the direct summand $\mathbb{R}$ in (\ref{DS}).
Thus, the assignment $f \mapsto \tau f+h$ defines an apimorphism
\begin{equation}\label{nu}
\nu: {\widetilde{F}}{}^0_{-1}\rightarrow
H^{(1)}(\delta|D)\,,\qquad \mathrm{Ker}\,\nu=\mathbb{R} \,.
\end{equation}
By definition, the variables $x^i$, $\bar\eta_i$ are characterized
by nonnegative ghost numbers and the same is true for the elements
of the space $\widetilde{F}^0_{-1}$.  The operator $\tau$
decreases the ghost number by one unit. Taking into account an
obvious correlation between the ghost number and momentum degree
as well as nilpotency of the odd variables $\bar\eta{}^i$,  we can
write
$$
H^{(1)}(\delta|D)=\bigoplus_{g=-1}^{n-1} H^g_{g+1}(\delta|D)\,.
$$

To further clarify the isomorphism  (\ref{KerD}) let us consider a
geometric interpretation of the space $\widetilde{F}^0_{-1}$.
Namely, we can think of functions
$$
a(x,\bar\eta)=\sum_{k=0}^\infty f(x)^{i_1\ldots
i_k}\bar\eta_{i_1}\ldots \bar \eta_{i_k}\in \widetilde{F}^0_{-1}
$$
as (inhomogeneous) polyvector fields on $\mathbb{R}^n$ with odd
variables $\bar\eta{}_i$ playing the role of the natural frame
$\partial/\partial x^i$. In this terms the exterior product of two
polyvector fields corresponds to the usual multiplication of
functions, while the Schouten bracket passes to
\begin{equation}\label{SB}
[a,b]= \frac{\partial a}{\partial \bar\eta_i}\frac{\partial
b}{\partial
x^i}-(-1)^{(\epsilon(a)+1)(\epsilon(b)+1)}\frac{\partial
b}{\partial \bar\eta_i}\frac{\partial a}{\partial x^i}\,.
\end{equation}
Both the multiplication operations are known to be compatible in
the sense of the graded Leibniz rule, so that we can speak of the
Gerstenhaber (or odd Poisson) algebra of  polyvector fields on
$\mathbb{R}^n$. We denote this algebra by
$\Lambda(\mathbb{R}^n)=\bigoplus\Lambda^p(\mathbb{R}^n)$. Now,
introducing the exterior ideal $J\subset \Lambda(\mathbb{R}^n)$
generated by $0$-vectors $T_a$ and $1$-vectors
$R_\alpha^i\bar\eta_i$, we can reformulate the involutivity
conditions (\ref{INV}) in the following way:
$$
[J,J]\subset J\,,\qquad [V,J]\subset J\,.
$$
The first relation just says that the exterior ideal $J$ is closed
for the Schouten bracket and so defines an ideal of the
Gerstenhaber algebra $\Lambda(\mathbb{R}^n)$. According to the
second relation this ideal is invariant with respect to the drift
vector field $V$. Define the $V$-invariant stabilizer of $J$ in
$\Lambda(\mathbb{R}^n)$ as
$$
\Lambda^J(\mathbb{R}^n)=\big\{ a\in \Lambda(\mathbb{R}^n)\;|\;
[V,a]\subset J\,,\; [a,J]\subset J \big \}\,.
$$
The space $\Lambda^J(\mathbb{R}^n)$ is clearly a subalgebra of
$\Lambda(\mathbb{R}^n)$ containing $J$ and  we can introduce the
quotient Gerstenhaber algebra
$\Lambda_J(\mathbb{R}^n)=\Lambda^J(\mathbb{R}^n)/J$. Now the
defining relation  (\ref{PrCon}) for the relative $\delta$-cocycle
$\tau f+h$ implies that $f$ represents an element of
$\Lambda^J(\mathbb{R}^n)$,
$$
   [V,f]=T_ah^a+h^\alpha R_\alpha\,,\qquad [T_a, f]= T_bh_a^b +h_a^\alpha R_\alpha\,,\qquad [R_\alpha,
    f]=T_ah^a_\alpha+h^\beta_\alpha R_\beta\,.
$$
This leads us to the following identification of the relative
homology groups belonging to $\mathrm{Ker} D_s^\ast$:
\begin{equation}\label{H=L}
H^{(1)}_0(\delta|D)\simeq\Lambda^0_J(\mathbb{R}^n)/\mathbb{R}\,,\qquad
H^{(1)}_m(\delta |D)\simeq \Lambda_J^m(\mathbb{R}^n)\,, \qquad
m=1,\ldots, n\,.
\end{equation}
Notice that all these groups are nested in resolution degree 1.
The general results on the local BRST cohomology obtained in
\cite{KLS2}, \cite{KLS1} suggest the following physical
interpretation of the groups (\ref{H=L}).  The space
$\Lambda^0_J(\mathbb{R}^n)$ is identified with the space of
conservation laws, then the quotient\footnote{Any constant is
obviously an integral of motion, but its gradient gives the zero
characteristic.} $\Lambda^0_J(\mathbb{R}^n)/\mathbb{R}$ coincides
with the space of characteristics of the system (\ref{INF}). The
subalgebra $\Lambda_J^1(\mathbb{R}^n)$ is naturally identified
with the Lie algebra of global symmetries. The space
$\Lambda^2_J(\mathbb{R}^n)$ is isomorphic, by definition, to the
space of nontrivial Lagrange structures. In more detail these
Lagrange structures will be discussed in Sec. 6, where we shall
identify them with the so-called weak Poisson brackets. Here, we
only mention that a Lagrange structure $P\in
\Lambda^2_J(\mathbb{R}^n)$ is called \textit{integrable} if
$[P,P]=0\in \Lambda^3_J(\mathbb{R}^n)$. This allows us to regard
$\Lambda^3_J(\mathbb{R}^n)$ as the space of potential obstructions
to  integrability of  Lagrange structures. In case $n=2$, each
Lagrange structure appears to be integrable for dimensional
reasons. As for the groups (\ref{H=L}) with momentum degree $>2$,
their interpretation as ``the spaces of'' or ``obstructions to''
is obscure to us at present.

It remains to consider the cokernel of the operator $D^\ast_s$.
Unfortunately, the description of the space
$\mathrm{Coker}D^\ast_s$ appears to be less explicit than the
kernel space. We know that each element $f\in
\widetilde{F}^{0}_{s+1}$ is a $\delta$-cycle and thus a cycle of
$\delta$ modulo $D$. From (\ref{Df}) it then follows that the
relative cycles of the form $f=D_s'g+Lg $ span  the image of
$D^\ast_s$. Therefore
\begin{equation}\label{cokerD}
    \mathrm{Coker}D^{\ast}_s=\frac{\widetilde{F}^{0}_{s+1}}{I_{s+1}\cup\mathrm{Im}(L+D'_s)}\,,\qquad H^{(0)}(\delta|D)\simeq
    \lim_{\rightarrow}\mathrm{Coker}D^{\ast}_s\,.
\end{equation}
Notice that all the elements of $\mathrm{Coker}D^\ast_s$ are
nested in zero resolution degree.

To gain greater insight into what the groups (\ref{cokerD}) are
about, consider  a mechanical system  without gauge symmetries and
constraints, that is, a system of ordinary differential equations
associated to the vector field $V$. Then the algebra
$\widetilde{F}{}^0_s=\widetilde{F}^0_{-1}$ can be identified with
$\Lambda(\mathbb{R}^n)$, the ideal $I_s$ is absent, and the
operator $L+D'_s$ reduces to the commutator with the drift $V$.
Therefore, $H^{(0)}(\delta|D)\simeq \Lambda(\mathbb{R}^n)/[V,
\Lambda(\mathbb{R}^n)]$. Now suppose $V$ is a vector field that
vanishes at $x_0\in \mathbb{R}^n$ together with its first partial
derivatives. Then the Schouten bracket $[V,W]$ vanishes at the
point $x_0$, too, for any polyvector field $W$. Therefore each
polyvector field that does not vanish at $x_0$ represents a
nontrivial class of the $\delta$ modulo $D$ cohomology. This
demonstrates nontriviality  of the  group  $H^{(0)}(\delta|D)$
even for systems with trivial phase-space topology. On the other
hand, if the vector field $V$ can be rectified  in the whole of
$\mathbb{R}^n$ (and so has no stationary points), then
$H^{(0)}(\delta|D)=0$. Indeed, in rectifying coordinates
$V=\partial/\partial x^1$ and for any polyvector field $W$ on
$\mathbb{R}^n$ we have the representation $W=[V,\widetilde{W}]$
with $\widetilde{W}=\int dx^1 W$.

\subsection{The group $H(\gamma, H(\delta|D))$} We begin
with a simple remark  that in resolution degree zero any relative
$\delta$-cycle is necessarily an ``absolute'' one as there is no
total derivatives of resolution degree minus one. This implies the
isomorphism
\begin{equation}\label{GCH}
H(\gamma, H(\delta|D))\simeq H(\gamma|D^\ast, H(\delta)) \,,
\end{equation}
using which we can set $\displaystyle H(\gamma,
H(\delta|D))=\lim_{\rightarrow} H(\gamma|D_s^\ast, H(F_s))$. Since
all the $\delta$-homology concentrates in resolution degree $0$,
so does the relative cohomology of $\gamma$, that is, $H(\gamma,
H^{(1)}(\delta|D))=0$.  Let $\gamma_s$ denote the differential in
$H(F_s)$ induced by the action of $\gamma$ in $F_s$ and let
$H(\gamma_s)=\bigoplus H^g(\gamma_s)$ denote the corresponding
cohomology group. Now to describe the relative $\gamma$-cohomology
group (\ref{GCH}) we can apply the mapping cone construction to
the cochain transformations $D_s^\ast: H(F_s)\rightarrow
H(F_{s+1})$ in perfect analogy to our computation of the relative
$\delta$-homology. This time, however, it is convenient to combine
the ``$\gamma$-counterparts'' of the exact triangle diagrams
(\ref{D1}) and (\ref{D2}) into a singe diagram, which looks like:
\begin{equation}\label{DIAGR} \xymatrix{H(\gamma, H({F}{}^0_0))\simeq
H(\gamma_s)\quad\ar[rr]^{\widetilde{D}_s}& & \quad H(\gamma_{s+1})
\ar[dl]\simeq H(\gamma, H({F}{}^0_0))\\
&H(\mathrm{Con} D^\ast_s)\ar[ul]^{\beta}{} \ar[dl]& \\
H(\mathrm{Coker} D^\ast_s)\ar[rr]\quad & &\quad
H(\mathrm{Ker}D^\ast_s)\ar[ul]_\alpha}
\end{equation}
Here $\widetilde{D}{}_s$ is the operator induced by $D^\ast_s$ on
cohomology. The group we are interested in is given by the lower
left corner of the diagram. In principle, it can be computed from
the bottom triangle provided we know the groups
$H(\mathrm{Ker}D^\ast_s)$ and $H(\mathrm{Con} D^\ast_s)$. By
definition, the group $\mathrm{Con}(D^\ast_s)$ is given by the
direct  sum $H(F_s)\oplus H(F_{s+1})$ endowed with the action of
the coboundary operator $\gamma$:
$$
\gamma(a,b)=(-\gamma_s a, D^\ast_s a+\gamma_{s+1} b)\,,\qquad
\mathrm{gh}(a,b)=\mathrm{gh}\,b=\mathrm{gh}\, a-1\,.
$$
The homomorphisms $\alpha$ and $\beta$ are induced by the natural
imbedding $a\mapsto (a, 0)$ and the natural projection
$(a,b)\mapsto a$. It follows from the top triangle in
(\ref{DIAGR}) that
$$
H(\mathrm{Con} D^\ast_s)\simeq \mathrm{Ker} \widetilde{D}_s\oplus
\mathrm{Coker} \widetilde{D}_s\,.
$$
To compute the kernel and cokernel of the operator
$\widetilde{D}_s$ consider the identity
\begin{equation}\label{D=K}
D f=Kf+\delta(\tau f)+[\gamma, \rho] f\,,
\end{equation}
which holds for any $f\in F_0^0$; here the operators $K$ and
$\rho$ are give by
$$
    K=-V^{i}\frac{\partial }{\partial
    x^{i}}+\bar{\eta}_{j}\frac{\partial V^{j}}{\partial x^{i}}\frac{\partial }{\partial
    \bar\eta_i}+
    c^{\beta}E_{\beta}^{\alpha}\frac{\partial }{\partial c^\alpha}+
    \left(c^{\alpha}F_{\alpha}^{ia}\bar{\eta}_{i}+\bar{\xi}^{b}D_{b}^{a}\right)\frac{\partial }{\partial
    \bar{\xi}^a}\,,\qquad\rho =\bar{\eta}^a\frac{\partial }{\partial \bar{\xi}^a}-\lambda^\alpha\frac{\partial }{\partial
c^\alpha}\,.
    $$
Since the operator $\rho$ anti-commutes with $\delta$, we can
interpret equality (\ref{D=K}) by saying that when restricted to
$\gamma$-cocycles from $\widetilde{F}^0_0$ the action of $D$
coincides with that of $K$ modulo $\gamma$-coboundaries.
Furthermore, as one can easily verify, the operator $K$ leaves
invariant the spaces $\widetilde{F}{}_0^0$, $I_0$ and $I_0\cup
\mathrm{Im}\gamma_0$, inducing thus a well-defined operator
$\widetilde{K}$ in the quotient space $\gamma_0^{-1}I_0/(I_0\cup
\mathrm{Im}\gamma_0)$. By virtue of the isomorphism (\ref{ISO-G})
we can identify the spaces $\mathrm{Ker}\widetilde{D}{}_s$ and
$\mathrm{Ker} \widetilde{K}$. Explicitly,
$$
\mathrm{Ker}{\widetilde{D}_s}\simeq
\mathrm{Ker}\widetilde{K}\simeq\frac{\gamma_0^{-1}I_0\cap
K^{-1}(I_0\cup \mathrm{Im}\gamma_0)}{I_0\cup
\mathrm{Im}\gamma_0}\,.
$$
For the kernel of $\widetilde{D}_s$ we have the following
representation:
$$
\mathrm{Coker}{\widetilde{D}_s}\simeq
\mathrm{Coker}\widetilde{K}\simeq\frac{\gamma_0^{-1}I_0}{I_0\cup
\mathrm{Im}\gamma_0\cup K\gamma^{-1}_0I_0}\,.
$$
With this result on the cohomology of the mapping cone, we can now
turn to the study of the group $H(\mathrm{Coker} D^\ast_s)$
entering the bottom triangle of diagram (\ref{DIAGR}). There are
two observations about this exact triangle: (i) the operator
$\gamma$ induces the zero differential in $\mathrm{Ker} D^\ast_s$,
so that $H(\mathrm{Ker}D^\ast_s)\simeq \mathrm{Ker} D^\ast_s$, and
(ii) the homomorphism $\alpha$ is an injection. The first fact is
easily seen by comparing the action of the operators $L_0$ and
$\gamma_0$ in $\widetilde{F}^0_{-1}$, while the second assertion
follows immediately from injectivity of the composition
$\beta\alpha$. The details are left to the reader. The
homomorphism $\alpha$ being injective, the bottom triangle in
(\ref{DIAGR}) reduces to the short exact sequence
\begin{equation*} \xymatrix{0
\ar[r]&\mathrm{Ker}D^\ast_s \ar[r]& \mathrm{Ker}
\widetilde{D}_s\oplus \mathrm{Coker} \widetilde{D}_s\ar[r] &
H(\mathrm{Coker}D^\ast_s) \ar[r]&0}\,,
\end{equation*}
where we made use of the established isomorphisms. Taking into
account that $\mathrm{Im}\alpha\subset
\mathrm{Ker}\widetilde{D}_s$, we can write
\begin{equation}\label{REL-G}
H(\mathrm{Coker}D^\ast_s)\simeq \frac{\mathrm{Ker}
\widetilde{D}_s} {\mathrm{Ker}D^\ast_s} \bigoplus
\mathrm{Coker}\widetilde{D}_s \,.
\end{equation}
The quotient in the right hand side can be understood as follows.
By (\ref{D=K}), a $\gamma$-cocycle $f\in \widetilde{F}{}^0_0$
gives rise to an element from $\mathrm{Ker}\widetilde{D}_s$ iff
$Kf=\gamma h+g$ for some $h\in {\widetilde{F}}^0_0$ and $g\in
I_0$. Read from right to left Eq. (\ref{D=K}) says that any such
$f$ defines the relative $\gamma$-cocycle $\rho f+ h$.  Among the
elements of  $\mathrm{Ker}\widetilde{D}_s$ are the ``absolute''
$\gamma$-cocycles from $\widetilde{F}^0_{-1}$.  Being independent
of $c$'s and $\bar\xi$'s, these $\gamma$-cocycles are all
annihilated by the operator $\rho$ and so do not contribute to the
relative $\gamma$-cohomology. This explains the structure of the
first summand in (\ref{REL-G}).

The above consideration can now be summarized in the following
formula:
$$
H(\gamma, H(\delta|D))\simeq \frac{\gamma_0^{-1}I_0\cap
K^{-1}(I_0\cup \mathrm{Im}\gamma_0)}{I_0\cup \mathrm{Im}\gamma_0
\cup (\gamma_0^{-1}I_0\cap\widetilde{F}_{-1}^0)}\bigoplus
\frac{\gamma_0^{-1}I_0}{I_0\cup \mathrm{Im}\gamma_0\cup
K\gamma^{-1}_0I_0}\,.
$$

We close this section with an explicit  example clarifying the
geometric origin of the (relative) $\gamma$-cohomology. Consider
the following system of differential algebraic equations:
\begin{equation}\label{eq-m}
\dot x-\lambda y=0\,,\qquad \dot y+\lambda x=0\,,\qquad
x^2+y^2-1=0\,.
\end{equation}
Comparing these equations with the general form of an involutive
system (\ref{INF}), it is easy to see that the constraint surface
$\Sigma$ is given here by the unit circle standardly imbedded  in
$xy$-plane and the gauge distribution is spanned by the single
vector field $ R=x{\partial_y}- y{\partial_x}$ generating
rotations. One can also check that the system meets both the
involutivity (\ref{INV}) and full rank (\ref{IRRED}) conditions.
By (\ref{GSYM}) and (\ref{NI}) each variable $\lambda$ results in
a gauge symmetry, and each constraint gives rise to a Noether
identity. In the case at hand, the gauge transformations are given
by
\begin{equation}\label{gtr}
\delta_{\varepsilon} x=\varepsilon y\,,\qquad \delta_{\varepsilon
}y=-\varepsilon x\,,\qquad \delta_\varepsilon
\lambda=\dot\varepsilon\,,
\end{equation}
while the Noether identity has the form
$$
2x(\dot x-\lambda y)+2y(\dot y+\lambda x)-D(x^2+y^2-1)=0\,.
$$
The gauge orbits foliate the plane onto concentric circles, one of
which coincides with the constraint surface. The physical phase
space, being isomorphic to the quotient $\Sigma/\sim$, is given by
a point, so that the system possesses no physical degrees of
freedom. As a result, the space of local physical observables is
exhausted by constant functions, $H^0(s_0)\simeq H^0(\gamma,
H(\delta))\simeq \mathbb{R}$. However, as we shall see in a
moment, there are nontrivial physical observables with values in
local functionals. Whereas the constants are physically observable
by definition (the ground field), the presence of nonlocal
observables is not a common property shared by all dynamical
systems. The classical BRST charge associated to our system reads
$$
\begin{array}{rl}
 \displaystyle\Omega_1=\int dt&\!\!\!\Big\{\bar\eta_x(\dot x-\lambda
y)+\bar\eta_y(\dot y+\lambda x)+\bar\eta(x^2+y^2-1)\\[4mm]
&+ c(y\bar x-x\bar y+\eta_y\bar\eta_x-\eta_x\bar\eta_y-\dot{\bar
\lambda})+\bar\xi(x\eta_x+y\eta_y-\frac12\dot\eta) \Big\}
\end{array}
$$
As there are no higher structure functions, the classical BRST
differential is mere the sum of the Koszul-Tate and longitudinal
differentials,
$$s_0=\delta+\gamma\,,\qquad
s_0^2=\delta^2=\gamma^2=0\,.$$ Now one can easily see that the
gauge ghost $c$ is BRST invariant,
$$
\delta c=0\,,\quad\gamma c=0 \quad \Rightarrow\quad s_0 c=0\,,
$$
and the corresponding classes of $s_0$- and $\gamma$-cohomology
are nontrivial. Indeed, if they were trivial, there would exist a
smooth function $f$ of $x$ and $y$ such that
\begin{equation}\label{cfg}
c=\gamma f+\delta g
\end{equation}
for some $g$. Let us introduce the polar coordinate system
$(r,\varphi)$ instead of the Cartesian coordinates  $(x,y)$. Then
in a collar neighborhood of the constraint surface $r=1$, the
function $f$ can be regarded as a smooth function of $r$ and
$\varphi$ such that $f(r,\varphi+2\pi)=f(r,\varphi)$. The
generator of the gauge distribution takes the form
$R=\partial_{\varphi}$.  Equation (\ref{cfg}) implies that
$$
(\partial_\varphi f)(1,\varphi)=1\,,
$$
whatever the function $g$. But the last equality  is impossible as
the derivative of a periodic function must vanish at least at two
points.

Applying the operator $\rho$ yields a nontrivial class of the
relative $\gamma$-cohomology, namely, $ \lambda=-\rho c$. The BRST
invariance of the integral $\Lambda=\int \lambda dt$ amounts to
its gauge invariance provided that the gauge parameter obeys the
zero boundary conditions:
$$s_0\Lambda=\int \dot c
dt=0\quad \Leftrightarrow\quad \delta_{\varepsilon}\Lambda=\int
\dot\varepsilon dt=0\,.$$ The gauge invariance of the functional
$\Lambda$ admits also a purely geometric explanation. Let us treat
$x$ and $y$ as $0$-forms and $\lambda$ as a $1$-form on the time
interval. Using the equations of motion (\ref{eq-m}), we can bring
the gauge transformations (\ref{gtr}) to the form of infinitesimal
reparametrizations:
$$
\delta_{\tilde{\varepsilon}}x=\dot x \tilde{\varepsilon}\,,\qquad
\delta_{\tilde{\varepsilon}}y=\dot y \tilde{\varepsilon}\,,\qquad
\delta_{\tilde{\varepsilon}}\lambda=D(\tilde{\varepsilon}\lambda)\,,\qquad
\varepsilon ={\tilde{\varepsilon}}\lambda\,.
$$
Then the functional $\Lambda$ is given by the integral of  the
$1$-form $\lambda$ over an interval (a one-dimensional manifold
with boundary), and hence it is invariant under diffeomorphisms.
Thus, we are lead to conclude that not only do the BRST cohomology
groups carry some valuable information about the physical sector
of the theory, but they also `feel' a particular realization of
the physical phase space by means of imbedding and/or
factorization.

\section{The total BRST charge}

The classical BRST charge, as its name suggests, incorporates all
the ingredients of the classical theory: the equations of motion,
their gauge symmetries and Noether identities. The corresponding
BRST complex provides concise and rigorous definitions for such
important notions of classical dynamics as physical observables,
rigid symmetries, and conservation laws. Whereas the classical
equations of motion are enough to formulate the classical dynamics
they are certainly insufficient for constructing a
quantum-mechanical description of the system. Any quantization
procedure has to involve one or another additional
geometric/algebraic structure. Within the path-integral
quantization, for instance, it is the action functional that plays
the role of such an additional structure. The procedure of
canonical quantization relies on the Hamiltonian form
of dynamics, involving a non-degenerate Poisson bracket and a
Hamiltonian. Either approach assumes the existence of a
variational formulation for the classical equations of motion (the
least action principle), and becomes inapplicable beyond the scope
of variational dynamics. The extension of these quantization
methods to general non-variational systems was proposed in
\cite{LS0}, \cite{KazLS}. In both the cases the structure
responsible for quantization is obtained as the deformation of the
corresponding classical BRST differential in the category of
$L_\infty$-algebras; in so doing, the classical BRST differential
is identified with the first structure map $L_1$. For the most
part, the quantum properties of the theory are determined  by the
second structure map $L_2$, that is, the first order deformation
of the classical BRST differential. In the Hamiltonian picture of
dynamics, or still better the phase-space approach,  $L_2$ is
identified with a weak Poisson structure \cite{LS0}, while in the
Lagrangian or covariant approach it is known as a Lagrange
structure \cite{KazLS}. It goes without saying that different
choices for the deformation of classical BRST differential can
generally result in different quantum theories.

The aim of this and the next two sections is to explain a
relationship between the two mentioned approaches to quantization
of non-variational gauge systems in the case of mechanical systems
brought to the involutive normal form (\ref{INF}). To begin with
we recall the definition of the total BRST charge.

Just as the path-integral quantization of Lagrangian gauge
theories is formulated by means of a master action on the
ghost-extended configuration space of fields, so the covariant
quantization of non-variational theories is defined in terms of a
single functional called the BRST charge. The latter can be viewed
as the deformation of the classical BRST charge $\Omega_1$ by
terms of higher momentum degree,
\begin{equation}\label{OMEGA}
\Omega=\Omega_1+\sum_{p=2}^\infty \Omega_p\,,\qquad
\mathrm{Deg}\,\Omega_p=p\,.
\end{equation}
The only condition on the deformation (besides being local,
Grassmann odd, and of ghost number 1) is  that the total BRST
charge $\Omega$ obeys the same master equation as the classical
one, i.e.,
\begin{equation}\label{MASTER EQUATION}
\{\,\Omega,\,\Omega\,\}=0\,.
\end{equation}
On substituting the expansion (\ref{OMEGA}) into (\ref{MASTER
EQUATION}), we get  the infinite chain  of equations
\begin{equation}\label{MASTER EQUATION EXPANSION}
\{\,\Omega_1,\,\Omega_1\,\}=0,\qquad
\{\,\Omega_1,\,\Omega_2\}=0,\qquad
\{\,\Omega_2,\,\Omega_2\,\}=2\{\,\Omega_1,\,\Omega_3\,\}\,,\qquad
\ldots\,.
\end{equation}
The first equation is automatically satisfied  for the classical
BRST charge $\Omega_1$. Then the second equation identifies the
leading term of the deformation, $\Omega_2$, as a relative cocycle
of the classical BRST differential $s_0=\{\,\Omega_1,\,\cdot\,\}$.
The deformation is called {\textit{regular}} if $[\Omega_2]\neq 0\in
H^{1}_{2}(s_0|D)$ and \textit{trivial} if $\Omega$ is canonically
equivalent to $\Omega_1$. In the latter case there exists an even
local functional $G$ of ghost number zero such that
$$
    \Omega=e^{\{G,\,\cdot\,\}}\Omega_1\,,\qquad
    \mathrm{Deg}\,G\geq 2\,.
$$
As the canonically equivalent systems are physically
indistinguishable, we can confine ourselves to considering only
nontrivial deformations. In our previous paper \cite{KLS2} the
following alternative was proven: every deformation of the
classical BRST charge associated to a mechanical system is either
regular or trivial. Non-triviality of the class $[\Omega_2]$ is,
of course, only a necessarily condition for the existence of a
nontrivial deformations starting with $\Omega_2$. As is usual in
deformation theory, the necessarily and sufficient condition for
the existence of a regular deformation  is that  all the Massey
powers of $[\Omega_2]$ can be made zero simultaneously
\cite{KLS2}. Indeed, due to the Jacoby identity the Poisson square
of the cocycle $\Omega_2$ is annihilated by $s_0$, and hence, we
have the class $[\{\Omega_2,\Omega_2\}]\in H^2_3(s_0|D)$. This
class, denoted usually by $\langle [\Omega_2],[\Omega_2]\rangle$,
is known as the Massey square of $[\Omega_2]$. One can easily see
that the Massey square depends actually on the class $[\Omega_2]$,
and not on its particular representative $\Omega_2$. For a general
discussion of the Mossey products in the category of graded Lie
algebras we refer the reader to \cite{R}, \cite{FL}. The Poisson
bracket $\{\Omega_2,\Omega_2\}$, representing the Massey square,
enters the left hand side of the third equation in (\ref{MASTER
EQUATION EXPANSION}). Since the the right hand side of the
equation is proportional to the coboundary $s_0\Omega_3$, we are
lead to conclude that the second order deformation $\Omega_2$ of
the classical BRST charge extends to the third order iff
$\langle[\Omega_2],[\Omega_2]\rangle=0$. Actually, it is also the
necessary and sufficient condition for the existence of the total
BRST charge $\Omega$. The reason is that all the higher Massey
powers
$$\langle
[\underbrace{\Omega_2],[\Omega_2],\ldots,[\Omega_2}_{m}]\rangle\,,\quad
m=3,4,\ldots\,,
$$ belong to the groups $ H^2_{m+1}(s_0|D)$ which are known to vanish
for mechanical systems \cite{KLS2}.

The total BRST charge admits also an interesting algebraic
interpretation, which gives a further elucidating glimpse into the
nature of regular deformations and their relation to the basic
ingredients of the Batalin-Vilkovisky (BV) formalism \cite{HT}.
Let $\mathcal{A}$ denote the space of local functionals of
momentum degree zero. In \cite{KazLS}, it was observed that each
total BRST charge (\ref{OMEGA}) endows the space $\mathcal{A}$
with the structure of $L_\infty$-algebra \cite{LSt}. The
corresponding structure maps $L_n: \mathcal{A}^{\otimes
n}\rightarrow \mathcal{A}$ are defined through the derived bracket
construction \cite{Vor}:
\begin{equation}\label{Lnn}
L_n: a_1\otimes a_2\otimes\cdots \otimes a_n\quad \mapsto\quad
(a_1,a_2,\ldots, a_n)=\{\cdots\{\Omega_n,a_1\},a_2\},\cdots,
a_n\}\in \mathcal{A}\,.
\end{equation}
In particular, the first structure map is given simply by the
classical BRST differential $s_0$ and the second structure map
defines  the 2-bracket
\begin{equation}\label{2br}
(a,b)=\{\{\Omega_2, a\},b\}\qquad \forall a,b\in \mathcal{A}\,.
\end{equation}
By definition, the $2$-bracket is  Grassmann odd and graded
symmetric,
$$
(a,b)=(-1)^{\epsilon(a)\epsilon(b)}(b,a)\qquad \forall a,b\in
\mathcal{A}\,.
$$
As for the graded Jacobi identity, it is replaced by the following
relation:
\begin{equation}\label{wji}
((a,b),c)+
(-1)^{\epsilon(b)\epsilon(c)}((a,c),b)+(-1)^{\epsilon(a)(\epsilon(b)+\epsilon(c))}((a,b),c)=-\Delta(a,b,c)\,,
\end{equation}
where the trilinear  functional $\Delta$, describing deviation
from the standard Jacobi identity, is determined by the third
order term in the total BRST charge (\ref{OMEGA}), namely,
$$
\Delta(a,b,c)=s_0(a,b,c)+(s_0a,b,c)+(-1)^{\epsilon(a)\epsilon(b)}(a,s_0b,c)+(-1)^{(\epsilon(a)+\epsilon(b))\epsilon(c)}(a,b,s_0c)\,.
$$
Following the physical terminology, we call (\ref{2br}) the
\textit{weak anti-bracket}  and refer to (\ref{wji}) as the
\textit{weak Jacobi identity}. The second relation in (\ref{MASTER
EQUATION EXPANSION}) implies that the classical BRST differential
$s_0$ and the weak anti-bracket are compatible in the sense of the
graded Leibniz rule
\begin{equation}\label{LR}
s_0(a,b)=-(s_0a,b)-(-1)^{\epsilon(a)}(a,s_0b)\qquad \forall a,b\in
\mathcal{A}\,.
\end{equation}
As a consequence, the weak anti-bracket descends to the classical
BRST cohomology, inducing an odd Lie bracket in the space
$H_0(s_0)$.

In a particular case, where the expansion  (\ref{OMEGA}) for the
total BRST charge stops at the second term, i.e.,
$\Omega=\Omega_1+\Omega_2$, the bracket (\ref{2br}) enjoys all the
properties of the usual BV anti-bracket \cite{HT}, including the
Jacobi identity. If we further assume the anti-bracket  to be
non-degenerate, then the classical BRST differential is
necessarily given by an anti-Hamiltonian vector field
$s_0=(S,\,\cdot\,)$ generated by some BV master action $S\in
\mathcal{A}$. The latter obeys the classical master equation
$(S,S)=0$ by virtue of $s_0^2=0$.  This is the most concise,
though a somewhat formal,  explanation of how the standard BV
formalism for Lagrangian systems fits in this more general
quantization approach.

\section{The Lagrange structure and the weak Hamiltonian structure}
The discussion of the previous section can be summarized by saying
that the total BRST charge $\Omega$ of a mechanical system is
completely specified (up to canonical transform) by a classical
BRST charge $\Omega_1$ and a relative BRST cocycle $\Omega_2$
satisfying the only condition
\begin{equation}\label{MSq}
\langle [\Omega_2], [\Omega_2]\rangle=0\,.
\end{equation}
It is the condition which ensures that the weak anti-bracket
(\ref{2br}) in $\mathcal{A}$ induces a genuine anti-bracket (=odd
Lie bracket) in the cohomology space $H_0(s_0|D)$.

We are now going to examine equation (\ref{MSq}) more closely,
using our knowledge about the structure of the local BRST
cohomology associated to involutive systems of ODEs. Investigation
of this question will lead us eventually to establishing an
explicit one-to-one correspondence between the concepts of a
Lagrange structure \cite{KazLS} and a weak Hamiltonian structure
\cite{LS0} for this particular class of dynamical systems.

In Section 4.3, we have shown the existence of the short exact
sequence
$$
\xymatrix{0 \ar[r]&{\mathbb{R}}\ar[r]^\mu&
\Lambda_J(\mathbb{R}^n)\ar[r]^\nu&H^{(1)}(\delta|D) \ar[r]&0}\,,
$$
where the monomorphism $\mu$ is the natural inclusion and the
epimorphism $\nu$ is defined by Eq. (\ref{nu}). The space
$\Lambda_J(\mathbb{R}^n)$ carries the structure of a graded Lie
algebra with the Lie bracket induced by the Schouten bracket on
polyvector fields. Notice that the space $\mathrm{Im}\,\mu \simeq
\mathbb{R}$, being identified with the space of constant functions
on $\mathbb{R}^n$, belongs to the center of
$\Lambda_J(\mathbb{R}^n)$. This allows us to define the quotient
Lie algebra $\Lambda_J(\mathbb{R}^n)/\mathbb{R}$, whose carrier
vector space is, by definition, isomorphic to $H^{(1)}(\delta|D)$.
The push forward of the Lie bracket on
$\Lambda_J(\mathbb{R}^n)/\mathbb{R}$ by means of $\nu$ defines
then the Lie algebra structure on the cohomology space
$H^{(1)}(\delta|D)$. Namely, if $a$ and  $ b$ are two elements of
$H^{(1)}(\delta|D)$ such that $a=\nu(\alpha)$ and $b=\nu(\beta)$
for some $\alpha,\beta \in \Lambda_J(\mathbb{R}^n)$, then
\begin{equation}\label{LIE-BR}
    \{a, b\}=\nu([\alpha,\beta])\,.
\end{equation}
Here we deliberately denote the push forward Lie bracket on
$H^{(1)}(\delta|D)$ by braces. The reason is that the right hand
side of (\ref{LIE-BR}) exactly coincides with the cohomology class
of the Poisson bracket of relative $\delta$-cocycles representing
the classes $a$ and $b$. The last fact can  also be seen from the
following construction. As established in \cite{KLS2}, the group
$H^{(1)}(\delta|D)$ is isomorphic to the direct product
$\Pi=\bigoplus_{g=-1}^\infty H^g_{g+1}(s_0|D)$. The corresponding
isomorphism $\varkappa: H^{(1)}(\delta|D)\rightarrow \Pi $ is
defined in the following way. Each representative cocycle $a$ of a
class $[a]\in \Pi$ can be expanded according to the resolution
degree,
$$
a=a^{_{(1)}}+a^{_{(2)}}+a^{_{(3)}}+\cdots\,,\qquad \deg \,
a^{_{(r)}}=r\,.
$$
The leading term has resolution degree 1 and is annihilated by the
Koszul-Tate differential. By definition, we set
$\varkappa([a])=[a^{_{(1)}}]\in H^{(1)}(\delta|D)$. Since the
action of the classical BRST differential $s_0$ is Hamiltonian,
the Poisson bracket on the space of local functionals passes
through the cohomology making $\Pi$ into a graded Lie algebra. The
pull back of this Lie algebra structure via the isomorphism
$\varkappa$ gives the above Lie bracket (\ref{LIE-BR}) on
$H^{(1)}(\delta|D)$. Thus, we  arrive at the following commutative
diagram of the Lie algebra isomorphisms:
\begin{equation}\label{LIE-ISO}
\xymatrix{&H^{(1)}(\delta|D)\ar[dr]^{\varkappa}&\\
\Lambda_J(\mathbb{R}^n)/\ar[ru]^\nu\mathbb{R}\ar[rr]^{\varkappa\nu}&
& \Pi=&\hspace{-10mm}\displaystyle \bigoplus_{g=-1}^\infty
H^g_{g+1}(s_0|D)}
\end{equation}

Let us now come back to the regular deformation (\ref{OMEGA})
governed by the class $[\Omega_2]\in H^1_2(s_0|D)$. In view of the
comments above this class has uniquely defined preimages in
$\Lambda^2_J(\mathbb{R}^n)$ and $H^{(1)}_2(\delta|D)$:
\begin{equation}\label{OLP}
[\Omega_2]=\varkappa([L])=\varkappa\nu([P])\,.
\end{equation}
The element $[L] $ of $H^{(1)}_2(\delta|D)$ is known as the
\textit{Lagrange structure} \cite{KLS2}. We see that for the
involutive systems of ODEs,  each Lagrange structure defines (and
is defined by) a unique class $[P]$ of
$\Lambda^2_J(\mathbb{R}^n)$. By the definition of
$\Lambda^2_J(\mathbb{R}^n)$, the bivector field
$P=P^{ij}\partial_i\wedge\partial_j$, representing the class
$[P]$, obeys the relations
\begin{equation}\label{PVRT}
[T_a,P]=-Y^\alpha_a R_\alpha-T_b G^b_a\,,\qquad
[R_\alpha,P]=W_\alpha^\beta \wedge R_\beta- T_a M^a_\alpha
\,,\qquad [V,P]=Z^\alpha \wedge R_\alpha-T_aN^a
\end{equation}
for some polyvector fields $Y$, $G$, $W$, $M$, $Z$, $N$. Applying
the map (\ref{nu}), one can see that the corresponding Lagrange
structure $[L]=\nu([P])$ is represented by the relative
$\delta$-cocycle
\begin{equation}\label{LS}\begin{array}{rl}
L=&2P^{ij}\bar{x}_i\bar{\eta}_j+\eta^k\partial_k
P^{ij}\bar{\eta}_i\bar{\eta}_j\\[5mm]&
-Z^{\alpha i}\bar\lambda_\alpha \bar{\eta}_i +\eta_a
N^{aij}\bar{\eta}_i\bar{\eta}_j+Y_a^\alpha
\bar{\lambda}_\alpha\bar{\eta}^a + \eta_b
G_a^{bi}\bar{\eta}^a\bar{\eta}_i- \lambda^\alpha W_\alpha^{\beta
i}\bar{\lambda}_\beta\bar{\eta_i}+ \lambda^\alpha \eta_a
M_\alpha^{aij}\bar{\eta}_i\bar{\eta}_j\,.\end{array}
\end{equation}
This cocycle incorporates all the polyvector fields entering the
right hand sides of the structure relations (\ref{PVRT}). For the
mechanical systems without gauge symmetries and constraints these
structure relations are absent and the corresponding Lagrange
structure is determined by the first line in (\ref{LS}).

Due to the Lie algebra isomorphisms (\ref{LIE-ISO}) and the
identifications (\ref{OLP}) the following conditions are pairwise
equivalent:
\begin{equation}\label{MMM}
    \langle [\Omega_2],[\Omega_2]\rangle=0\quad \Leftrightarrow\quad \langle
    [L],[L]\rangle=0 \quad \Leftrightarrow \quad \langle
    [P],[P]\rangle=0\,.
\end{equation}
In \cite{KLS2}, a Lagrange structure was called
\textit{integrable} if all its Massey powers can be made zero. For
mechanical systems this integrability condition boils down to
vanishing of the Massey square of $[L]$. Relation (\ref{MMM})
tells us that in the case of involutive systems of ODEs one can
make one step further and reduce verification of the middle
equality in (\ref{MMM}) to verification of the rightmost one. This
is an added reason for working with involutive normal forms, since
the structure of a representative  $P$ is generally much simpler
than that of $L$, as is seen from (\ref{LS}). In terms of
representatives, the vanishing of the Massey square of
$[P]=\nu^{-1}([L])$  amounts to the condition $[P,P]\in J$ or,
explicitly,
\begin{equation}\label{WJI}
[P,P]=U^\alpha \wedge R_\alpha-T_a S^a
\end{equation}
for some vector fields $U^\alpha$ and bivector fields $S^a$.

A bivector field $P\in \Lambda^2(\mathbb{R}^n)$ is said to define a
\textit{weak Poisson structure} on $\mathbb{R}^n$ if it satisfies
the first two relations in (\ref{PVRT}) together with (\ref{WJI}).
Another name for $P$ is $P_\infty$-structure \cite{Vor}. Relation
(\ref{WJI}) is called the \textit{weak Jacobi identity}. Given a
weak Poisson structure $P$, a vector field $V$ is called
\textit{weakly Hamiltonian} if it obeys the third relation in
(\ref{PVRT}). The set of four polyvector fields $(V,R,T, P)$ is
referred to as a \textit{weak Hamiltonian structure} on
$\mathbb{R}^n$. If the right hand sides of relations (\ref{PVRT})
and (\ref{WJI}) are equal to zero, then the adjective ``weak'' can
be omitted. In this case, $P$ is just a Poisson bivector,  $V$ and
$R_\alpha$'s are the corresponding Poisson vector fields, and
$T_a$'s are Casimir functions for $P$. This is always true for
mechanical systems without gauge symmetries and constraints.

As we have seen in Sec. 4.2, the commutative algebra of physical
observables $H^0_0(s_0)$ with values in local functions is
isomorphic to a certain subquotient $\mathcal{F}$ of  the algebra
$C^\infty(\mathbb{R}^n)$. Namely, let $I$ denote the ideal of
$C^\infty(\mathbb{R}^n)$ generated by the functions $T_a$. In view
of the invulutivity conditions (\ref{INV}) the gauge distribution
$R$ preserves $I$ in the sense that $[R,I]\subset I$, and hence
its action descends to the quotient $C^\infty(\mathbb{R}^n)/I$. By
definition, the algebra $\mathcal{F}$ is constituted by the
$R$-invariant elements of $C^\infty(\mathbb{R}^n)/I$, cf.
(\ref{gadb}). In other words, a function $O\in
C^\infty(\mathbb{R}^n)$ represents an observable $[O]\in
\mathcal{F}\subset C^\infty(\mathbb{R}^n)/I$ if $[R_\alpha, O]\in
I$ and two such functions $O$ and $O'$ represent the same
observable, $[O]=[O']$, if $O-O'\in I$.

The weak Poisson structure $[P]\in \Lambda_J^2(\mathbb{R}^n)$
makes the commutative algebra $\mathcal{F}$ into a Poisson
algebra. The corresponding Poisson bracket is defined as a derived
bracket \cite{Vor} on representatives:
$$
\{[O_1],[O_2]\}_P=[[[P,O_1],O_2]]\qquad \forall [O_1], [O_2]\in
\mathcal{F}\,.
$$
Using the property of the Schouten bracket, one can easily verify
that this bracket operation is well-defined and enjoys all the
properties of a Poisson bracket: bilinearity, skew-symmetry, and
the Jacobi identity. Furthermore, the Poisson algebra
$\mathcal{F}$ comes equipped with a derivation naturally induced
by the drift $V$.   Equating this derivation to the time
derivative, we get the differential equation
$$
D[O]=[[V, O]]
$$
governing the evolution of a physical observable $ [O]\in
\mathcal{F}$. The nontrivial integrals of motion of the system
(\ref{INF}) correspond then to the $V$-invariant observables. They
constitute a Poisson subalgebra in $\mathcal{F}$, which, as a
linear space, is isomorphic to the space of conservation laws
$\Lambda_J^0(\mathbb{R}^n)\subset \mathcal{F}$. The last fact
 follows immediately from the definition of the space
$\Lambda_J^0(\mathbb{R}^n)$.

In the absence of quantum anomalies, the Poisson algebra
$(\mathcal{F}, \{\,\cdot\,,\,\cdot\,\})$ was shown to admit a
consistent deformation quantization  by means of a superextension
of Kontsevich's formality theorem \cite{LS0}, \cite{CaFe}. The
result of the deformation quantization is an associative
$\ast$-product in the space of quantum observables
$\mathcal{F}\otimes \mathbb{C}[[\hbar]]$ together with a
$\ast$-product derivation $\widehat{V}$ generating a one-parameter
family of automorphisms of the quantum algebra
($\mathcal{F}\otimes \mathbb{C}[[\hbar]], \ast$).

\section{Superfield formulation of the total BRST charge}

The weak Hamiltonian structure discussed in the previous section
admits a nice BRST description in terms of generating functions
\cite{LS0}. Let us briefly recall its main details. Given  a weakly
Hamiltonian system $(V,R,T,P)$, the phase space $\mathbb{R}^n$ of
coordinates $x^i$ is extended by the odd variables $\eta_a$ and
$c^\alpha$ called the ghosts: one $\eta$ for each constraint $T$ and
one $c$ for each gauge symmetry generator $R$. Denoting all the
variables by $\phi^A=(x^i, \eta_a, c^\alpha)$, one then redoubles
their number by introducing the dual variables
$\stackrel{_{\ast}}{\phi}_B=\{\stackrel{\ast}{x}_i,\stackrel{\ast}{\eta}{\!}^a,\stackrel{\ast}{c}_\alpha\}$
with opposite Garassmann parities. These are also called ghosts. The
superspace $W$ coordinatized by $\phi^A$ and
$\stackrel{\ast}{\phi}_A$ is endowed with the canonical
antisymplectic structure defined by the following  antibrackets (odd
Poisson brackets):
$$
    (\phi^A,\phi^B)=0\,,\qquad (\stackrel{_{\ast}}\phi_A,\phi^B)=\delta_A^B\,,
    \qquad
    (\stackrel{_{\ast}}{\phi}_A,\stackrel{_{\ast}}{\phi}_B)=0\,.
$$
Besides the Grassmann parity, all the variables carry three
additional $\mathbb{Z}$-gradings, which are called, respectively,
the ghost number, resolution degree and momentum
degree\footnote{In \cite{LS0}, the momentum degree was referred to
as the polyvector degree.}:
\begin{equation}\label{GRADINGS2}
\begin{array}{c}
\mathrm{gh}(x^i)=0\,,\qquad\mathrm{gh}(\eta^a)=-1\,,\qquad
\mathrm{gh}(c^\alpha)=1\,,\qquad
\mathrm{gh}(\stackrel{\ast}{\phi}_A)=
1-\mathrm{gh}({\phi}^A)\,,\\[3mm]
\mathrm{deg}(x^i)=\mathrm{deg}(\stackrel{_{\ast}}{x}{\!\!}_i)=\mathrm{deg}
(\stackrel{_{\ast}}{\eta}{\!\!}^a)=\mathrm{deg}(c^\alpha)=0\,,\qquad\mathrm{deg}(\stackrel{_{\ast}}{c}{}_{\alpha})=\mathrm{deg}({\eta}{}_{a})=1\,,\\[3mm]
\mathrm{Deg}(\phi^A)=0\,,\qquad\mathrm{Deg}(\stackrel{_{\ast}}{\phi}_B)=1\,.
\end{array}
\end{equation}
In the absence of fermionic degrees of freedom (all $x$'s are
even) the Grassmann parity and the ghost number are compatible in
the usual sense:
$$
\epsilon(\phi^A)=\mathrm{gh}(\phi^A)\,,\qquad
\epsilon(\stackrel{_{\ast}}{\phi}_B)=\mathrm{gh}(\stackrel{_{\ast}}{\phi}_B)\qquad
(\mathrm{mod} \;2)\,.
$$
Now all the structure relations  associated to the weak
Hamiltonian structure $(V,R,T,P)$ are compactly encoded in the
pair of master  equations
\begin{equation}\label{mE}
    (S, S)=0,\qquad (S,\Gamma )=0\,,
\end{equation}
where the generating functions $S$ and $\Gamma$ are subject to the
following grading and boundary conditions:
$$
\begin{array}{lll}
\mathrm{gh}({S})=2\,,&\qquad \epsilon(S)=0\,,&\qquad
\mathrm{Deg}(S)>0\,,\\[3mm]
\mathrm{gh}({\Gamma})=1\,,&\qquad\epsilon(\Gamma)=1\,,&\qquad
\mathrm{Deg}({\Gamma})>0\,,
\end{array}
$$

$$
S=\stackrel{_{\ast}}{\eta}{}^{a}T_{a}(x)+\stackrel{_{\ast}}{x}_{i}R^{i}_{\alpha}(x)c^{\alpha}+
\stackrel{_{\ast}}{x}_{i}\stackrel{_{\ast}}{x}_{j}P^{ij}(x)+\cdots\,,\qquad
\Gamma=\stackrel{_{\ast}}{x}_{i}V^{i}(x)+\cdots\,.
$$
The dots in the last line refer to the terms of positive
resolution degree. All these terms can be systematically found
from the master equations (\ref{mE}) by means of homological
perturbation theory with respect to the resolution degree
\cite{LS0}. As is seen, the bosonic function $S$ incorporates all
the ingredients of the weak Poisson structure: the phase-space
constraints $T$, the gauge symmetry generators $R$, and the weak
Poisson bivector $P$. The weakly Hamiltonian vector field $V$ --
the drift -- enters the fermionic function $\Gamma$. Expanding the
master equations (\ref{mE}) in powers of ghosts, one readily
recovers the involutivity conditions (\ref{INV}), defining
relations (\ref{PVRT}), (\ref{WJI}) for a weak Hamiltonian
structure, and the hierarchy of their differential consequences.

As with the total BRST charge $\Omega$, the generating function
$S$ gives rise to an $L_\infty$-structure on the space $A$ of
functions of momentum degree zero. If $S=\sum_{m=1}^\infty S_m$ is
the  expansion of $S$ with respect to the momentum degree, then
the $n$-th structure map $L_n:A^{\otimes n}\rightarrow A$ is given
by
\begin{equation}\label{Ln}
L_n: a_1\otimes a_2\otimes\cdots\otimes a_n\quad \mapsto\quad
\{a_1,a_2,\ldots,a_n\}= (\cdots (S_n,a_1),a_2),\cdots, a_n) \in
A\,.
\end{equation}
In particular, the second structure map defines the weak Poisson
bracket
\begin{equation}\label{WPB}
\{a,b\}=((S_2, a), b)=-(-1)^{\epsilon(a)\epsilon(b)}\{b,a\}
\end{equation}
satisfying the weak Jacobi identity
\begin{equation}\label{wji1}
\begin{array}{c} (-1)^{\epsilon(a)\epsilon(c)}\{\{a, b\}, c\} +
(-1)^{\epsilon(c)\epsilon(b)}\{\{c,
a\},b\}+(-1)^{\epsilon(b)\epsilon(a)}\{\{b, c\}, a\}\\[3mm]
=(-1)^{\epsilon(a)\epsilon(c)+1}\big(s_0(a,b,c)+(s_0a,b,c)+(-1)^{\epsilon(a)}(a,s_0b,c)+(-1)^{\epsilon(a)+\epsilon(b)}(a,b,s_0c)\big)
\end{array}
\end{equation}
where $s_0=(S_1,\,\cdot\,)$ is the classical BRST differential.
The operator $s_0$ differentiates the weak Poisson bracket
(\ref{WPB}) by the graded Leibniz rule. If $S_n=0$ for all $n>2$,
then the first master equation (\ref{mE}) implies that (\ref{WPB})
is a usual Poisson bracket determined by the Poisson bivector $P$.
For a nondegenerate  $P$ the classical BRST differential is given
then (locally) by a Hamiltonian vector fields $s_0=\{\Omega,
\,\cdot\,\}$, with $\Omega$ being the usual BFV-BRST-charge
\cite{HT}.

\vspace{3mm} \noindent \textit{Remark.} Formulae
(\ref{Lnn})-(\ref{LR}) and (\ref{Ln})-(\ref{wji1}) show a striking
algebraic parallelism in the two BRST formalisms for non-variational
systems. Notice, however, a difference in the symmetry properties of
the multibrackets (\ref{Lnn}) and (\ref{Ln}): the multibrackets
associated to the total BRST charge $\Omega$ are graded symmetric,
while those associated to the function $S$ are graded
skew-symmetric. Actually, there are two equivalent definitions of an
$L_\infty$-algebra, in terms of symmetric and skew-symmetric
multibrackets, and one may use either of them. Equivalence is
established by the parity reversion functor, see \cite[Remark
2.1]{Vor} for details.

\vspace{3mm}

In the previous section, we have shown that any Lagrange structure
compatible with an involutive system of ODEs defines and is defined
by some weakly Hamiltonian structure. So, there is a perfect
correspondence between both the pictures of one and the same
dynamics.  Our argumentation, however, was somewhat indirect and
heavily relied on the structure of local BRST cohomology. Below, we
are going to present a direct construction of the total BRST charge
$\Omega$ by the generating functions $S$ and $\Gamma$ of a weakly
Hamiltonian structure.  To that end, we shall follow the elegant
superfield approach proposed in quite a similar context by Damgaard
and Grigoriev \cite{GD} (see also \cite{BG2011}).

Consider the superspace $\mathbb{R}^{1|1}$ with one even
coordinate $t$, identified with time, and one odd coordinate
$\theta$, the odd superpartner of $t$. The smooth maps from
$\mathbb{R}^{1|1}$ to the antisymplectic space $W$ are described
by the superfields $\phi^A(t,\theta)$ and
$\stackrel{\ast}{\phi}_A(t,\theta)$, which form an infinite
dimensional superspace $\mathcal{W}$. The canonical antisymplectic
structure on $W$ induces then a canonical symplectic structure on
$\mathcal{W}$. The latter is defined by the Poisson brackets
\begin{equation}\label{PB3}
    \{\phi^A(z),\phi^B(z')\}=0\,,\qquad
    \{\stackrel{\ast}{\phi}_A(z),\phi^B(z')\}=\delta_A^B\delta(z-z')\,,\qquad
    \{\stackrel{\ast}{\phi}_A(z),\stackrel{\ast}{\phi}_B(z')\}=0\,,
\end{equation}
where $z=(t,\theta)$. Since $\theta^2=0$, each superfield contains
a pair of component fields that are just functions of time:
$$
\phi^A(t,\theta)=\phi^A_0(t)+\theta\phi^A_1(t)\,,\qquad
\stackrel{\ast}{\phi}_A(t,\theta)=\stackrel{\ast}{\phi}{\!\!}_A^0(t)
+\theta\stackrel{\ast}{\phi}{\!\!}_A^1(t)\,.
$$
If we set $\mathrm{gh}(\theta)=1$ and $\mathrm{Deg}(\theta)=0$,
then the ghost number and momentum degree of the component fields
are unambiguously determined by those of superfields
(\ref{GRADINGS2}). It is, however, imposable to prescribe a
definite resolution degree to $\theta$. The zero-components of
superfields define a trajectory in the antisymplectic space $W$.
Introducing the individual notation for their superpartners
$$
\phi^A_1(t)=\{\eta^i(t), -\xi_a(t), -\lambda^\alpha(t)\}\,,\qquad
\stackrel{\ast}{\phi}{}_A^1(t)=\{\bar x_i(t), \bar\eta {}^a(t),
\bar c_\alpha(t)\}
$$
and making identifications
$$\stackrel{\ast}{\phi}{\!}^0_A(t)=\{\stackrel{\ast}{x}_i(t),\stackrel{\ast}{\eta}{\!}^a(t),\stackrel{\ast}{c}_\alpha(t)\}
=\{\bar\eta_i(t),\bar\xi{}^a(t),\bar\lambda_\alpha(t)\}\,,
$$
we see that the set of component fields $\{\phi^A_0,
\phi^A_1,\stackrel{\ast}{\phi}{}_B^0,\stackrel{\ast}{\phi}{}_B^1\}$
exactly coincides with the set of fields
$\{\varphi^I,\bar\varphi_J\}$ from Sec. 2, including the
distribution of the Grassmann parities, ghost numbers and momentum
degrees. Furthermore, evaluating the Poisson brackets (\ref{PB3})
for the component fields, one can find that they are identical to
the Poisson brackets (\ref{PB}). This means that the infinite
dimensional symplectic superspaces $V$ and $\mathcal{W}$ with the
Poisson brackets (\ref{PB}) and (\ref{PB3}) are actually
isomorphic to each other.

Now we define an odd homomorphism $h$ relating the antibracket on
$W$ with the Poisson bracket on $\mathcal{W}$. To any function
$F(\phi,\stackrel{\ast}{\phi})$ on $W$ the homomorphism $h$
assigns the local functional
$$
h(F)=\int dtd\theta
F(\phi(t,\theta),\stackrel{\ast}{\phi}(t,\theta))\,.
$$
It is easy to check that $h$ is indeed a homomorphism of Lie
algebras, i.e.,
$$
\{h(F),h(G)\}=h((F,G))\,.
$$
The last property holds true even if one allows the functions $F$
and $G$ to depend on $t$ and $\theta$ as parameters. We are going
to apply this homomorphism  to the function
$$Q(\phi,\stackrel{\ast}{\phi},\theta)=S(\phi,\stackrel{\ast}{\phi})+\theta
\Gamma(\phi,\stackrel{\ast}{\phi})\,,$$ which is just a linear
combination of the generating functions of weak Hamiltonian
structure. It is clear that $\epsilon(Q)=0$ and
$\mathrm{gh}(Q)=1$. Regarding $\theta$ as an external odd
parameter, one can see that the master equations (\ref{mE}) are
equivalent to the single equation
\begin{equation}\label{QQ}
(Q,Q)=0\,.
\end{equation}
Consider now the functional
\begin{equation}\label{itd}
    \Omega=\int dt d\theta \Big(\stackrel{\ast}{\phi}_AD\phi^A+
    {Q}(\phi(t,\theta),\stackrel{\ast}{\phi}(t,\theta),\theta)\Big)\,,\qquad D\equiv-\theta\frac{\partial}{\partial
    t}\,.
\end{equation}
It satisfies all the grading conditions for the total BRST charge
and verification of the master equation yields
$$
\{\Omega,\Omega\}=h((Q,Q))+2\int dtd\theta DS =2\int
dt\frac{d}{dt}S(\phi_0(t),\stackrel{\ast}{\phi}{}^0(t))=0\,.
$$
Here we used the master equation (\ref{QQ}), the obvious identity
$D^2=0$, and the zero boundary conditions for the fields of
positive momentum degree.

Integration by $\theta$ in (\ref{itd}) yields  the total BRST
charge as the functional of component fields:
\begin{equation}\label{Q}
\Omega=\int dt\Big\{
(-1)^{\epsilon({\phi^A_0})+1}\stackrel{\ast}{\phi}{\!}^0_A\dot\phi^A_0+{\phi}_1^A\frac{\partial
S}{\partial
\phi^A}(\phi_0,\stackrel{\ast}{\phi}{\!}^0)+\stackrel{\ast}{\phi}{\!}_A^1\frac{\partial
S}{\partial
{\phi}^\ast_A}(\phi_0,\stackrel{\ast}{\phi}{\!}^0)+\Gamma(\phi_0,\stackrel{\ast}{\phi}{\!}^0)\Big\}\,.
\end{equation}
Expanding the last expression further in powers of ghosts, one can
see that the functional $\Omega$ meets also the boundary condition
for the total BRST charge associated with the involutive equations
(\ref{INF}) and the compatible Lagrange structure (\ref{LS}).
Thus, formula (\ref{itd}) establishes a desired correspondence
between the generating functions of a weak Hamiltonian structure
and the total BRST charge. Let us mention two special properties
of the BRST charge (\ref{Q}). First, the functional (\ref{Q})
involves no more than the first derivatives of fields, and these
derivatives enter the $\Omega$ in a vary peculiar way. Second, the
functional (\ref{Q}) is at most linear in $\phi_1^A$ and
$\stackrel{\ast}{\phi}{\!\!}_A^1$. The existence of such a
solution to the master $\{\Omega,\Omega\}=0$ is not easily seen
without resort to the superfield approach.

\section{Conclusion}

In this paper, we presented a detailed analysis of the local BRST
cohomology for general mechanical systems brought to the
involutive normal form. The term ``general''  means that (i) we do
not restrict ourselves to Lagrangian or constrained Hamiltonian
systems and (ii) any regular system of ODEs can be equivalently
reformulated in the involutive form at the cost of introducing
auxiliary variables. Starting from the involutive normal form, we
describe all the relevant groups of local BRST cohomology listed
at the end of Sec. 3. In particular, we have identified the groups
$H^{(1)}(\delta|D)$ with certain subquotients (\ref{H=L}) of the
algebra of polyvector fields on the phase space of the system.
Thus, an explicit evaluation of these groups for a given model
reduces to the standard problem of differential geometry. The most
notable homogeneous subgroups of $H^{(1)}(\delta|D)$ are those
associated with the spaces of conservation laws, global symmetries
and Lagrange structure. Using the results of Sec. 4, we establish
a one-to-one correspondence between the spaces of integrable
Lagrange structures and weakly Hamiltonian structures.
Establishing of such a correspondence is a matter of principle; it
is as fundamental for the general dynamics as the correspondence
between the BV and BFV quantization methods in the particular case
of variational systems. Although our consideration was restricted
to the mechanical systems, we hope that the computational
technique developed in this paper can also be used in field theory
with a due account of space locality. Finally, we gave a direct
superfield construction of the total BRST charge $\Omega$ by the
generating functions of the weakly Hamiltonian structure, Eqs.
(\ref{itd}), (\ref{Q}).  This generalizes the construction of Ref.
\cite{GD} for the BV master action in terms of the BRST charge and
unitarizing Hamiltonian. In the view of the aforementioned
correspondence between the Lagrange and weakly Hamiltonian
structures, it is natural to ask about the inverse construction of
the generating functions $S$ and $\Gamma$ by the total BRST charge
$\Omega$. Such a construction exists indeed, and we are going to
present it elsewhere.


\begin{thebibliography}{30}


\bibitem{HT}  M. Henneaux and C. Teitelboim, {Quantization of Gauge
Systems} (Princeton U.P., NJ, 1992).

\bibitem{AKSZ} M. Alexandrov, M. Kontsevich, A. Schwarz and O. Zaboronsky,
\textit{The Geometry of the Master Equation and Topological Quantum
Field Theory}, Int. J. Mod. Phys. \textbf{A 12} (1997) 1405-1430.


\bibitem{BF} I.A. Batalin and E.S. Fradkin, \textit{Operatorial quantizaion of dynamical systems subject to constraints.
A Further study of the construction}, Annales Poincare Phys. Theor.
\textbf{49}, No 2 (1988) 145-214.

\bibitem{BH} G. Barnich and M. Henneaux, \textit{Isomorphisms between the Batalin-Vilkovisky antibracket and the Poisson bracket},
J. Math. Phys. \textbf{37} (1996) 5273-5296.


\bibitem{GD} M.A. Grigoriev and P.H. Damgaard, \textit{Superfield BRST
charge and the master action},  Phys. Lett. \textbf{B474} (2000)
323-330.

\bibitem{LS0}S.L. Lyakhovich and A.A. Sharapov, \textit{BRST theory without Hamiltonian and Lagrangian},
JHEP \textbf{0503}:011.

\bibitem{KazLS} P.O. Kazinski, S.L. Lyakhovich and A.A.Sharapov,
\textit{Lagrange structure and quantization},  JHEP
\textbf{0507}:076.


\bibitem{CaFe} A. Cattaneo  and G. Felder, \textit{Relative formality
theorem and quantization of coisotropic submanifolds}, Adv. Math.
\textbf{208} (2007) 521-548.%10


\bibitem{LS} S.L. Lyakhovich and  A.A. Sharapov, \textit{Normal Forms and Gauge
Symmetries of Local Dynamics}, J. Math. Phys. \textbf{50} (2009)
083510.


\bibitem{LS1} S.L. Lyakhovich and A.A. Sharapov,
\textit{Schwinger-Dyson equation for non-Lagrangian field theory},
JHEP \textbf{0602}:007.

\bibitem{LS2} S.L. Lyakhovich and A.A. Sharapov,  \textit{Quantizing non-Lagrangian gauge theories: An augmentation method},
JHEP \textbf{0701}:047.


\bibitem{BBH1} G. Barnich, F. Brandt and M. Henneaux, \textit{Local BRST
cohomology in the antifield formalism. I. General theorems},
Commun. Math. Phys. \textbf{174} (1995) 57-91.

\bibitem{BBH2}G. Barnich, F. Brandt and M. Henneaux, \textit{Local BRST
cohomology in the antifield formalism. II. Application to
Yang-Mills theory}, Commun. Math. Phys. \textbf{174} (1995)
93-116.

\bibitem{BBH} G. Barnich, F. Brandt and M. Henneaux, \textit{Local BRST cohomology in gauge theories}, Phys. Rept. \textbf{338} (2000)
439-569.

\bibitem{KLS2}D.S. Kaparulin, S.L. Lyakhovich and A.A. Sharapov,
\textit{Local BRST cohomology in (non-)Lagrangian field theory},
JHEP \textbf{1109}:006.


\bibitem{K-S} Y. Kosmann-Schwarzbach, The Noether theorems: Invariance and
conservation laws in the twentieth century (Sources and Studies in
the History of Mathematics and Physical Sciences, Springer-Verlag,
2010).

\bibitem{KLS1} D.S. Kaparulin, S.L. Lyakhovich and A.A. Sharapov, \textit{Rigid symmetries and conservation laws in non-Lagrangian
field theory}, J. Math. Phys. \textbf{51} (2010) 082902.%6


\bibitem{BG1} G. Barnich and M. Grigoriev, \textit{A Poincare lemma for sigma models of AKSZ type}, J. Geom. Phys. \textbf{61} (2011)
663-674.

\bibitem{HT1988} M. Henneaux and C.Teitelboim, \textit{BRST
cohomology in Classical Mechanics}, Commun. Math. Phys. \textbf{115}
(1988) 213.

\bibitem{Seiler} W.M. Seiler, Involution, Algorithms and
Computation in Mathematics 24 (Springer-Verlag, Berlin-Heidelberg,
2010).%2

\bibitem{AS} A.A. Agrachev and Yu.L. Sachkov, Control theory from the geometric
viewpoint (Springer-Verlag, Berlin-Heidelberg, 2004).%4


\bibitem{Mac} S. Maclane, Homology (Springer-Verlag,
Berin-G\"ottingen-Heidelberg, 1963).%8


\bibitem{R} V. Retakh, \textit{Lie-Massey brackets and $n$-homotopically
multiplicative maps of differential graded Lie algebras}, J. Pure
and Appl. Algebra \textbf{89} (1993) 217-229.%12

\bibitem{FL} D. Fuchs and L. Lang Weldon, \textit{Massey brackets and
deformations}, J. Pure and Appl. Algebra \textbf{156} (2001)
215-229.%13

\bibitem{LSt} T. Lada and J. Stasheff, \textit{Introduction to sh Lie
algebras for physicists}, Int. J. Theor. Phys. \textbf{32} (1993)
1087-1103.%14

\bibitem{Vor} Th. Voronov, \textit{Higher derived brackets and homotopy
algebras}, J. Pure and Appl. Algebra \textbf{202} (2005) 133-153.%11

\bibitem{BG2011} G.Barnich and M.Grigoriev, \textit{First order parent
formulation for generic gauge field theories}, JHEP
\textbf{1101}:122.

\end{thebibliography}
\end{document}